\documentclass{emulateapj}

\usepackage{graphicx}
\usepackage{subfigure}
\usepackage{verbatim}
\usepackage{natbib}
\newcommand{\htwo}{H$_2$}
\newcommand{\HI}{H{\small I}}
\newcommand{\HII}{H{\small II}}

\newcommand{\acet}{C$_2$H$_2$}
\newcommand{\hcn}{HCN}
\newcommand{\cotwo}{CO$_2$}
\newcommand{\water}{H$_2$O}
\newcommand{\meth}{CH$_4$}

\newcommand{\neII}{[\ion{Ne}{2}]}
\newcommand{\neIII}{[\ion{Ne}{3}]}
\newcommand{\sIII}{[\ion{S}{3}]}
\newcommand{\sIV}{[\ion{S}{4}]}
\newcommand{\arIII}{[\ion{Ar}{3}]}
\newcommand{\clII}{[\ion{Cl}{2}]}
\newcommand{\neV}{[\ion{Ne}{5}]}
\newcommand{\feII}{[\ion{Fe}{2}]}
\newcommand{\oIV}{[\ion{O}{4}]}
\newcommand{\siII}{[\ion{Si}{2}]}
\newcommand{\arII}{[\ion{Ar}{2}]}
\newcommand{\um}{$\mu$m}

\shorttitle{MID-IR SPECTROSCOPY OF OH MEGAMASERS: PAPER I}
\shortauthors{WILLETT ET AL.}

\begin{document}

\title{Mid-infrared properties of OH megamaser host galaxies. I. \textit{Spitzer} IRS low- and high-resolution spectroscopy}

\author{Kyle W. Willett\altaffilmark{1}, Jeremy Darling\altaffilmark{1}, Henrik W. W. Spoon\altaffilmark{2}, Vassilis Charmandaris\altaffilmark{3,4}, \& Lee Armus\altaffilmark{5}}

\altaffiltext{1}{Center for Astrophysics and Space Astronomy, Department of Astrophysical and Planetary Sciences, UCB 391, University of Colorado, Boulder, CO 80309-0391; willettk@colorado.edu}
\altaffiltext{2}{Astronomy Department, Cornell University, Ithaca, NY 14853}
\altaffiltext{3}{University of Crete, Department of Physics and Institute of Theoretical \& Computational Physics, GR-71003, Heraklion, Greece}
\altaffiltext{4}{IESL/Foundation for Research and Technology-Hellas, GR-71110, Heraklion, Greece; and Chercheur Associ\'e, Observatoire de Paris, F-75014, Paris, France}
\altaffiltext{5}{Spitzer Science Center, California Institute of Technology, Pasadena, CA 91125}


\begin{abstract}
We present mid-infrared spectra and photometry from the Infrared Spectrograph on the \textit{Spitzer Space Telescope} for 51 OH megamasers (OHMs), along with 15 galaxies confirmed to have no megamaser emission above $L_{OH}=10^{2.3} L_\sun$. The majority of galaxies display moderate-to-deep 9.7~\um~amorphous silicate absorption, with OHM galaxies showing stronger average absorption and steeper 20--30~\um~continuum emission than non-masing galaxies. Emission from multiple polycyclic aromatic hydrocarbons (PAHs), especially at 6.2, 7.7, and 11.3~\um, is detected in almost all systems. Fine-structure atomic emission (including \neII, \neIII, \sIII, and \sIV) and multiple~\htwo~rotational transitions are observed in more than 90\% of the sample. A subset of galaxies show emission from rarer atomic lines, such as \neV, \oIV, and \feII. 50\% of the OHMs show absorption from water ice and hydrogenated amorphous carbon grains (HACs), while absorption features from \cotwo, \hcn, \acet, and crystalline silicates are also seen in several OHMs. Column densities of OH derived from 34.6~\um~OH absorption are similar to those derived from 1667~MHz OH absorption in non-masing galaxies, indicating that the abundance of masing molecules is similar for both samples. This data paper presents full mid-infrared spectra for each galaxy, along with measurements of line fluxes and equivalent widths, absorption feature depths, and spectral indices. 
\end{abstract}



\section{Introduction}
\label{sec-intro}

OH megamasers (OHMs) are 18-cm masers with integrated line luminosities on the order of $10^1-10^4$ L$_\sun$. They are an extremely rare phenomenon in the local universe, with roughly one hundred currently known out to a redshift of $z=0.265$ \citep{baa92}. All OHMs, including the more powerful ``gigamasers" ($L_{OH} > 10^4$ L$_\sun$) are associated with starburst nuclei in merging galaxies. OHMs have been identified in many different types of nuclear environments as classified by optical spectra, but the merging galaxies are without exception (ultra)luminous infrared galaxies ([U]LIRGs). Since OHMs are signposts of gas-rich merging galaxies, their presence can also indicate the existence of associated phenomena including massive black hole mergers and highly obscured circumnuclear starbursts \citep{dar07}. OHMs are a powerful tool in this respect due in large part to their ability to be seen at cosmic distances. In order to employ OHMs as tracers, however, the assumption must be made that the OH line properties remain constant as a function of cosmic time and host environment. An explanation of the physical mechanisms and conditions responsible for distinguishing OHMs from non-masing ULIRGs is thus vital for understanding both the megamaser phenomenon and the associated merger characteristics. 

Spectroscopic studies of the mid-IR emission in the host galaxies offer multiple diagnostics which can provide clues to the nature of the maser pumping mechanism and the associated OH emission. We used the Infrared Spectrograph (IRS) aboard the \textit{Spitzer Space Telescope} \citep{wer04} to study merging ULIRGs. Since the dusty nuclear regions are typically obscured at optical wavelengths, mid-IR observations can yield valuable information specific to the locations in which the OHMs are generated. These include measurements of the dust temperature and optical depth (from broadband photometry and absorption features), the excitation and temperature of the gas (molecular and fine-structure atomic lines), and high-ionization lines that can signal the presence of an active galactic nucleus (AGN), a possible heating source for the dust. 

This data paper presents full low- and high-resolution IRS spectra along with measured mid-IR properties for 51 OHMs and 15 non-masing ULIRGs. An accompanying paper (Willett et~al. 2011; Paper II) presents the full analysis, statistical comparisons of the masing and non-masing galaxies, and tests the viability of current OHM pumping models based on the IRS data. 


\section{The sample}\label{sec-sample}

The OHM host galaxies selected for IRS observations were primarily drawn from the Arecibo OHM survey \citep{dar02,dar02a}. We selected well-studied OHMs with unambiguous maser detections and large amounts of ancillary data (including OH line and radio continuum maps, near-IR imaging, and optical imaging and spectroscopy) to maximize scientific return on the sample. A lower threshold of $L_{OH}>10^{1.6}~L_\sun$ also eliminated extragalactic ``kilomasers'' from the sample, which are likely powered by different radiative processes than megamasers \citep{hen90}. 

In order to be detected in reasonable integration times using the IRS, we required that all potential targets have $S(60~\mu$m$)>0.8$~Jy as measured by the IRAS satellite. After removing objects already observed by the IRS \citep[largely through the GTO ULIRG program; {\it eg}][]{arm07}, we observed 24 galaxies in the redshift range $0.1<z<0.2$.

We supplemented these galaxies with additional spectra of OHM hosts publicly available through the {\it Spitzer} archive. To ensure uniformity of the data, we selected only galaxies from the archive that had full coverage with both the IRS low- and high-resolution modules. As of March 2008, the publicly available data from the archive yielded an additional 27 OHM galaxies (Table~\ref{tbl-summary}).  

\begin{deluxetable*}{llcclccrrl}
\tabletypesize{\scriptsize}
\tablecaption{Radio, optical, and FIR properties of OHMs and non-masing galaxies \label{tbl-summary}}
\tablewidth{0pt}
\tablehead{
\colhead{} & 
\colhead{IRAS FSC} & 
\colhead{RA} &
\colhead{Dec} & 
\colhead{z$_\sun$\tablenotemark{a}} &
\colhead{D$_L$} &
\colhead{log L$_{FIR}$\tablenotemark{b}} &
\colhead{log L$_{OH}$ \tablenotemark{c}} &
\colhead{$f_{1.4~GHz}$ \tablenotemark{d}} &
\colhead{Alt. desig.}
\\
\colhead{} & 
\colhead{} & 
\colhead{J2000.0} & 
\colhead{J2000.0} & 
\colhead{} & 
\colhead{[$h_{70}^{-1}$ Mpc]} &
\colhead{[$h_{70}^{-2} L_\sun$]} &
\colhead{[$h_{70}^{-2} L_\sun$]} &
\colhead{[mJy]} &
\colhead{}
}
\startdata
OHMs       & IRAS 01355$-$1814 & 01 37 57.4 & $-$17 59 21 & 0.191  & 929  & 12.49         &   2.75  &  $<5.0$&                \\  
           & IRAS 01418+1651   & 01 44 30.5 &   +17 06 05 & 0.0274 & 115  & 11.63         &   2.71  &   40.6 & III Zw 035     \\  
           & IRAS 01562+2528   & 01 59 02.6 &   +25 42 37 & 0.1658 & 788  & 12.19         &   3.31  &    6.3 &                \\  
           & IRAS 02524+2046   & 02 55 17.1 &   +20 58 43 & 0.1815 & 873  & 12.07$-$12.54 &   3.80  &    2.9 &                \\  
           & IRAS 03521+0028   & 03 54 42.2 &   +00 37 03 & 0.1522 & 718  & 12.59         &   2.49  &    6.7 &                \\  
           & IRAS 04121+0223   & 04 14 47.1 &   +02 30 36 & 0.1216 & 568  & 11.69$-$11.96 &   2.39  &    3.1 &                \\  
           & IRAS 04454$-$4838 & 04 46 49.5 & $-$48 33 33 & 0.0529 & 235  & 11.89         &   2.95  &  $<5.0$& ESO 203-IG 001 \\  
           & IRAS 06487+2208   & 06 51 45.8 &   +22 04 27 & 0.1437 & 678  & 12.34         &   2.87  &   10.8 &                \\  
           & IRAS 07163+0817   & 07 19 05.5 &   +08 12 07 & 0.1107 & 515  & 11.79         &   2.43  &    3.5 &                \\  
           & IRAS 07572+0533   & 07 59 57.2 &   +05 25 00 & 0.1894 & 926  & 12.31         &   2.80  &    5.0 &                \\  
           & IRAS 08201+2801   & 08 23 12.6 &   +27 51 40 & 0.1680 & 808  & 12.26         &   3.51  &   16.7 &                \\  
           & IRAS 08449+2332   & 08 47 51.0 &   +23 21 06 & 0.1510 & 723  & 12.05         &   2.65  &    6.1 &                \\  
           & IRAS 08474+1813   & 08 50 18.3 &   +18 02 01 & 0.1450 & 692  & 12.19         &   2.76  &    4.2 &                \\  
           & IRAS 09039+0503   & 09 06 34.2 &   +04 51 25 & 0.1250 & 589  & 12.16         &   2.88  &    6.6 &                \\  
           & IRAS 09539+0857   & 09 56 34.3 &   +08 43 06 & 0.1290 & 608  & 12.09         &   3.53  &    9.5 &                \\  
           & IRAS 10035+2740   & 10 06 26.3 &   +27 25 46 & 0.1662 & 794  & 12.26         &   2.55  &    6.3 &                \\  
           & IRAS 10039$-$3338 & 10 06 04.8 & $-$33 53 15 & 0.0341 & 154  & 11.74         &   2.98  &   24.7 &                \\  
           & IRAS 10173+0828   & 10 20 00.2 &   +08 13 34 & 0.0480 & 222  & 11.86         &   2.77  &   10.8 &                \\  
           & IRAS 10339+1548   & 10 36 37.9 &   +15 32 42 & 0.1965 & 969  & 12.35         &   2.71  &    5.1 &                \\  
           & IRAS 10378+1109   & 10 40 29.2 &   +10 53 18 & 0.1362 & 646  & 12.35         &   3.35  &    8.9 &                \\  
           & IRAS 10485$-$1447 & 10 51 03.1 & $-$15 03 22 & 0.1330 & 629  & 12.23         &   2.99  &    4.4 &                \\  
           & IRAS 11028+3130   & 11 05 37.5 &   +31 14 32 & 0.1990 & 975  & 12.39         &   3.03  &    5.0 &                \\  
           & IRAS 11180+1623   & 11 20 41.7 &   +16 06 57 & 0.1660 & 801  & 12.27         &   2.40  &    4.2 &                \\  
           & IRAS 11524+1058   & 11 55 02.8 &   +10 41 44 & 0.1784 & 868  & 12.19         &   3.04  &    5.0 &                \\  
           & IRAS 12018+1941   & 12 04 24.5 &   +19 25 10 & 0.1687 & 814  & 12.48         &   2.96  &    6.5 &                \\  
           & IRAS 12032+1707   & 12 05 47.7 &   +16 51 08 & 0.2170 & 1082 & 12.64         &   4.21  &   28.7 &                \\  
           & IRAS 12112+0305   & 12 13 46.0 &   +02 48 38 & 0.0730 & 335  & 12.38         &   3.04  &   23.8 &                \\  
           & IRAS 12540+5708   & 12 56 14.2 &   +56 52 25 & 0.0422 & 188  & 12.42         &   2.94  &  309.9 & Mrk 231        \\  
           & IRAS 13218+0552   & 13 24 19.9 &   +05 37 05 & 0.2051 & 1011 & 12.44         &   3.50  &    5.3 &                \\  
           & IRAS 13428+5608   & 13 44 42.1 &   +55 53 13 & 0.0378 & 167  & 12.18         &   2.61  &  145.4 & Mrk 273        \\  
           & IRAS 13451+1232   & 13 47 33.3 &   +12 17 24 & 0.1220 & 571  & 12.21         &   2.46  & 5398.0 & 4C +12.50      \\  
           & IRAS 14059+2000   & 14 08 18.7 &   +19 46 23 & 0.1237 & 580  & 11.94         &   3.40  &    7.5 &                \\  
           & IRAS 14070+0525   & 14 09 31.2 &   +05 11 32 & 0.2644 & 1346 & 12.87         &   4.50  &    5.2 &                \\  
           & IRAS 14553+1245   & 14 57 43.4 &   +12 33 16 & 0.1249 & 585  & 11.87         &   2.33  &    3.8 &                \\  
           & IRAS 15327+2340   & 15 34 57.1 &   +23 30 11 & 0.0181 & 80   & 12.22         &   2.65  &  326.8 & Arp 220       \\  
           & IRAS 16090$-$0139 & 16 11 40.5 & $-$01 47 06 & 0.1339 & 628  & 12.57         &   3.52  &   20.9 &                \\  
           & IRAS 16255+2801   & 16 27 38.1 &   +27 54 52 & 0.1340 & 627  & 11.94         &   2.62  &    5.0 &                \\  
           & IRAS 16300+1558   & 16 32 21.4 &   +15 51 45 & 0.2417 & 1212 & 12.80         &   2.91  &    7.9 &                \\  
           & IRAS 17207$-$0014 & 17 23 21.9 & $-$00 17 01 & 0.0428 & 188  & 12.45         &   3.10  &   82.4 &                \\  
           & IRAS 18368+3549   & 18 38 35.4 &   +35 52 20 & 0.1162 & 536  & 12.24         &   2.91  &   21.0 &                \\  
           & IRAS 18588+3517   & 19 00 41.2 &   +35 21 27 & 0.1067 & 489  & 11.92         &   2.58  &    5.9 &                \\  
           & IRAS 20100$-$4156 & 20 13 29.5 & $-$41 47 35 & 0.1296 & 603  & 12.68         &   4.13  &  $<5.0$&                \\  
           & IRAS 20286+1846   & 20 30 55.5 &   +18 56 46 & 0.1347 & 633  & 12.06         &   3.47  &    5.0 &                \\  
           & IRAS 21077+3358   & 21 09 49.0 &   +34 10 20 & 0.1764 & 846  & 12.10$-$12.24 &   3.32  &    9.4 &                \\  
           & IRAS 21272+2514   & 21 29 29.4 &   +25 27 50 & 0.1508 & 709  & 11.99$-$12.14 &   3.71  &    4.4 &                \\  
           & IRAS 22055+3024   & 22 07 49.7 &   +30 39 40 & 0.1269 & 587  & 12.19         &   2.79  &    6.4 &                \\  
           & IRAS 22116+0437   & 22 14 09.9 &   +04 52 24 & 0.1939 & 937  & 12.12$-$12.32 &   2.83  &    8.4 &                \\  
           & IRAS 22491$-$1808 & 22 51 49.2 & $-$17 52 23 & 0.0778 & 346  & 12.19         &   2.46  &    5.9 &                \\  
           & IRAS 23028+0725   & 23 05 20.4 &   +07 41 44 & 0.1496 & 701  & 11.86$-$12.06 &   3.34  &   19.5 &                \\  
           & IRAS 23233+0946   & 23 25 56.2 &   +10 02 49 & 0.1279 & 591  & 12.18         &   2.80  &   11.6 &                \\  
           & IRAS 23365+3604   & 23 39 01.3 &   +36 21 09 & 0.0645 & 283  & 12.19         &   2.52  &   28.7 &                \\  
           \hline                                                                                                                      
Non-masing & IRAS 00164$-$1039 & 00 18 50.4 & $-$10 22 08 & 0.0272 & 113  & 11.36         & $<1.25$ & $<5.0$ & Arp 256        \\  
           & IRAS 01572+0009   & 01 59 50.2 &   +00 23 41 & 0.1630 & 774  & 12.47         & $<2.12$ &  26.7  & Mrk 1014       \\  
           & IRAS 05083+7936   & 05 16 46.4 &   +79 40 13 & 0.0537 & 237  & 11.93         & $<1.94$ &  41.4  &                \\  
           & IRAS 06538+4628   & 06 57 34.4 &   +46 24 11 & 0.0214 & 93.6 & 11.24         & $<0.89$ &  64.3  & UGC 3608       \\  
           & IRAS 08559+1053   & 08 58 41.8 &   +10 41 22 & 0.1480 & 705  & 12.18         & $<1.72$ & $<5.0$ &                \\  
           & IRAS 09437+0317   & 09 46 20.6 &   +03 03 30 & 0.0205 & 93.5 & 11.15         & $<1.01$ & $<5.0$ &                \\  
           & IRAS 10565+2448   & 10 59 18.1 &   +24 32 34 & 0.0431 & 194  & 12.04         & $<1.66$ &  57.0  &                \\  
           & IRAS 11119+3257   & 11 14 38.9 &   +32 41 33 & 0.1890 & 923  & 12.48         & $<2.07$ & 110.4  &                \\  
           & IRAS 13349+2438   & 13 37 18.7 &   +24 23 03 & 0.1076 & 500  & 11.39         & $<1.72$ &  20.0  &                \\  
           & IRAS 15001+1433   & 15 02 31.9 &   +14 21 35 & 0.1627 & 781  & 12.42         & $<2.04$ &  16.9  &                \\  
           & IRAS 15206+3342   & 15 22 38.0 &   +33 31 36 & 0.1244 & 582  & 12.13         & $<1.75$ &  11.2  &                \\  
           & IRAS 20460+1925   & 20 48 17.3 &   +19 36 54 & 0.1807 & 868  & 12.03         & $<2.15$ &  18.9  &                \\  
           & IRAS 23007+0836   & 23 03 15.6 &   +08 52 26 & 0.0163 & 64.9 & 11.43         & $<0.63$ & 181.0  & NGC 7469       \\  
           & IRAS 23394$-$0353 & 23 42 00.8 & $-$03 36 55 & 0.0232 & 95.4 & 11.11         & $<1.18$ & $<5.0$ & Arp 295B       \\  
           & IRAS 23498+2423   & 23 52 26.0 &   +24 40 17 & 0.2120 & 1037 & 12.44         & $<2.25$ &   6.8  &                \\  
\enddata                                                                                                     
\tablenotetext{a}{Heliocentric optical redshift \citep{dar02}.}                                              
\tablenotetext{b}{Computed according to the prescription of \cite{san96}, with a scale factor of $C=1.6$. IRAS photometry is from \citet{san03}; a range in $L_{FIR}$ means that the object was not detected by IRAS at 100~\um.}
\tablenotetext{c}{OH fluxes are from \cite{dar02,dar02a}; limits are computed according to Eqn.~\ref{eqn-maxoh}.}
\tablenotetext{d}{Flux densities at 1.4~GHz are from the NRAO VLA Sky Survey \citep{con98}.}
\end{deluxetable*}

In order to provide a baseline for analysis of the OHMs, we also identified a control sample of ULIRGs that showed no megamaser emission above a firm limit. To identify these galaxies, we drew on non-detections from OH surveys by \cite{baa92, sta92, dar00, dar01, dar02} and \cite{ken02}. The upper limit for OH emission is conservatively derived from the rms noise in the spectrum at 1667~MHz, assuming a boxcar line profile with a linewidth $\Delta v=150$~km/s and a 1.5$\sigma$ detection: 

\begin{equation}
\label{eqn-maxoh}
L_{OH}^{max} = 4 \pi D_L^2 (1.5 \sigma) \left(\frac{\Delta v}{c}\right) \left(\frac{\nu_0}{1+z}\right)
\end{equation}

For this control sample, we set an upper limit of $L_{OH}^{max}<10^{2.3}~L_\sun$; this limit compromises between ensuring that all but the faintest megamaser emission is excluded and yielding a reasonable number of objects in the control sample for statistical analysis. All 51 OHMs in our IRS sample have $L_{OH}$ above this limit. 

In addition to selecting galaxies based on OH non-detection, we imposed two additional criteria to ensure that the control sample was as similar as possible to the OHM hosts. Firstly, we set a lower limit on the far-infrared luminosity \citep{san96} of the non-masing galaxies as measured by their IRAS fluxes. OHMs occur exclusively in IR-bright galaxies, due to the fact that the maser is pumped primarily by rotational transitions of a few hundred K above the ground state \citep{baa82,hen87}. \cite{dar02} show that the relationship between the OH and infrared luminosities is a power law with $L_{OH}\propto L_{FIR}^{1.2}$. Since no OHM observed with the IRS has $L_{FIR}<10^{11}L_\sun$, we established this as the lower limit for inclusion in the non-masing sample. 

\begin{figure}
\includegraphics[scale=0.5]{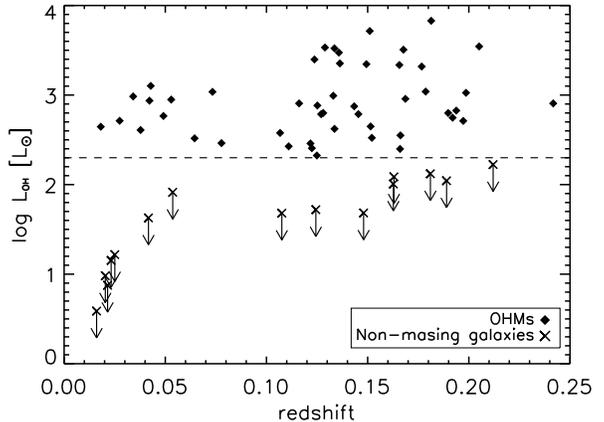}
\caption{Distribution of integrated OH luminosity for the IRS samples as a function of redshift. For non-masing galaxies, the OH luminosity is an upper limit calculated from Equation~\ref{eqn-maxoh}. The dashed line at $L_{OH}=10^{2.3}~L_\sun$ is the upper limit on possible OH emission for the non-masing control sample. \label{fig-ncutplot}}
\end{figure}

Secondly, a cutoff in redshift space is applied to sample a sufficiently large volume ($V\sim~1~\textrm{Gpc}^3$) in order to avoid systematic effects such as the Malmquist bias. The available data in the archive contained many more galaxies at lower redshifts ($z<0.05$) than those further away. To avoid over-weighting the control sample towards galaxies at low redshifts, we sorted galaxies that met the $L_{OH}^{max}$ and $L_{IR}$ criteria into bins of $\Delta~z=0.02$. For bins where the number of non-masing galaxies exceeded those of OHMs, we randomly removed objects from the non-masing bins until the numbers were equal. This reduced the control sample to 1 galaxy at $0<z<0.02$ and 4 galaxies at $0.02<z<0.04$; for all other redshift bins, no such adjustments were necessary. As of March 2008, there existed 15 suitable candidates in the \textit{Spitzer} archive qualifying for the non-masing control sample (Table~\ref{tbl-summary}). Figure~\ref{fig-ncutplot} shows the distribution of OH luminosity for all objects as a function of redshift; for the non-masing galaxies, we display upper limits as computed in Equation~\ref{eqn-maxoh}. 
 
Throughout this paper, we assume the WMAP5 cosmology with $H_0=70.5$, $\Omega_M=0.274$, and $\Omega_\Lambda=0.726$ \citep{hin09}. 

\section{Observations}\label{sec-obs}

We observed 24 OHMs with the IRS from August 2006 through December 2007. The IRS contains four modules in two different spectroscopic resolutions: low- (LR) and high-resolution (HR) \citep{hou04}. The short-high (SH) and long-high (LH) modules operate at a resolution of $R\sim 600$; the short-low (SL1 \& SL2) and long-low (LL1 \& LL2) modules operate at resolutions ranging from $R\sim56-127$, depending on the observing wavelength. Two orders, SL3 ($7.3-8.7$~\um) and LL3 ($19.4-21.7$~\um), cover the overlapping range between the first and second orders in both SL and LL. We used these only as checks for the absolute flux calibration between the different orders. 

All targets used the Staring Mode Astronomical Observing Template (AOT) with the galaxies placed at two nod positions approximately one-third and two-thirds the length of the slit. We observed targets in all six modules as well as an equal-time, off-target exposure in the LH module to be used for sky subtraction. 

Although the majority of our sources had no previous IRAS detections in either the 12 or 25~\um~bands, we extrapolated the likely flux based on the colors of ULIRGs at similar redshifts and the measured fluxes at 60~\um. We chose cycle times intended to yield signal-to-noise ratios of $S/N=50-100$ for the low-resolution modules and $S/N\ge10$ for the high-resolution modules, allowing for accurate measurement of faint emission and absorption features, as well as accurate spectral decomposition. 

For the 24 objects in the dedicated OHM program, we took dedicated sample-up-the-ramp (SUR) peakup observations in both the blue (16~\um) and red (22~\um) filters in sample-up-the-ramp (SUR) mode for photometric calibration of the spectra. The SUR peakups did not coincide in time with the spectroscopy for the majority of targets. None of the galaxies selected from the {\it Spitzer} archive possessed SUR peakup data; 16 archived galaxies have double correlated sampling (DCS) peakups with slightly worse photometric accuracy than those with SUR data. Only one galaxy, IRAS~10173+0828, had no peakup observations of any kind. Measured peakup fluxes are given in Table~\ref{tbl-puflux}. 

The \textit{Spitzer} beam is diffraction limited past 6~\um, with slit widths for the spectral modules between $3-11\arcsec$. The vast majority of the nuclei in merging ULIRGs have angular separations less than the instrument PSF, and thus can be treated as point sources. For the few galaxies close enough to be resolved, we chose staring mode observations centered on the IR-dominant nuclear region of the galaxy. The only available observations of the double nucleus in IRAS~10485$-$1447 were centered on the western nucleus. Details of the IRS observations are given in Table~\ref{tbl-obs}. 

\begin{deluxetable*}{lllccrrrrc}
\tabletypesize{\scriptsize}
\tablecaption{IRS observation log\label{tbl-obs}}
\tablewidth{0pt}
\tablehead{
\colhead{Object} & 
\colhead{Date} & 
\colhead{Peakup type} & 
\colhead{SL1} & 
\colhead{SL2} & 
\colhead{LL1} & 
\colhead{LL2} & 
\colhead{SH} & 
\colhead{LH} &
\colhead{Program}
}
\startdata
IRAS 01355$-$1814 &             & BPU-offset & $60 \times 2$ & $60 \times 2$ & $30 \times 2$ & $30 \times 2$ & $120 \times 4$ & 240 $\times$ 5  & 4,6   \\
IRAS 01418+1651   &             & BPU        & $14 \times 2$ & $14 \times 2$ & $14 \times 4$ & $14 \times 4$ & $120 \times 5$ & 60  $\times$ 7  & 2,3   \\
IRAS 01562+2528   & 2007 Sep 09 & BPU-offset & $14 \times 7$ & $14 \times 7$ & $30 \times 3$ & $30 \times 3$ & $120 \times 2$ & 60  $\times$ 2  & 1     \\
IRAS 02524+2046   & 2007 Sep 09 & BPU        & $14 \times 7$ & $14 \times 7$ & $30 \times 3$ & $30 \times 3$ & $120 \times 2$ & 60  $\times$ 2  & 1     \\
IRAS 03521+0028   &             & BPU-offset & $60 \times 2$ & $60 \times 2$ & $30 \times 3$ & $30 \times 3$ & $120 \times 3$ & 60  $\times$ 4  & 4     \\
IRAS 04121+0223   & 2007 Oct 05 & BPU-offset & $60 \times 2$ & $60 \times 2$ & $30 \times 3$ & $30 \times 3$ & $120 \times 2$ & 60  $\times$ 2  & 1     \\
IRAS 04454$-$4838 &             & BPU        & $60 \times 2$ & $60 \times 2$ & $30 \times 3$ & $30 \times 3$ & $120 \times 3$ & 240 $\times$ 2  & 3     \\
IRAS 06487+2208   & 2007 May 04 & BPU-offset & $14 \times 7$ & $14 \times 7$ & $30 \times 3$ & $30 \times 3$ & $120 \times 2$ & 60  $\times$ 2  & 1     \\
IRAS 07163+0817   & 2007 May 03 & BPU        & $14 \times 6$ & $14 \times 6$ & $30 \times 3$ & $30 \times 3$ & $120 \times 2$ & 60  $\times$ 2  & 1     \\
IRAS 07572+0533   & 2007 May 04 & BPU        & $14 \times 6$ & $14 \times 6$ & $30 \times 3$ & $30 \times 3$ & $120 \times 2$ & 60  $\times$ 2  & 1     \\
IRAS 08201+2801   & 2007 May 03 & BPU-offset & $14 \times 7$ & $14 \times 7$ & $30 \times 3$ & $30 \times 3$ & $120 \times 2$ & 60  $\times$ 2  & 1     \\
IRAS 08449+2332   & 2007 May 04 & BPU-offset & $14 \times 6$ & $14 \times 6$ & $30 \times 3$ & $30 \times 3$ & $120 \times 2$ & 60  $\times$ 2  & 1     \\
IRAS 08474+1813   & 2007 Dec 05 & BPU        & $14 \times 6$ & $14 \times 6$ & $30 \times 3$ & $30 \times 3$ & $120 \times 2$ & 60  $\times$ 2  & 1     \\
IRAS 09039+0503   &             & BPU        & $60 \times 2$ & $60 \times 2$ & $30 \times 4$ & $30 \times 4$ & $120 \times 4$ & 240 $\times$ 3  & 5,6   \\
IRAS 09539+0857   &             & BPU        & $60 \times 2$ & $60 \times 7$ & $30 \times 4$ & $30 \times 4$ & $120 \times 3$ & 240 $\times$ 3  & 5,6   \\
IRAS 10035+2740   & 2007 Jun 09 & BPU-offset & $14 \times 7$ & $14 \times 7$ & $30 \times 3$ & $30 \times 3$ & $120 \times 2$ & 60  $\times$ 2  & 1     \\
IRAS 10039$-$3338 &             & BPU        & $14 \times 6$ & $14 \times 6$ & $14 \times 4$ & $14 \times 4$ & $30  \times 6$ & 60  $\times$ 2  & 3     \\
IRAS 10173+0828   &             & none       & $60 \times 1$ & $60 \times 1$ & $14 \times 4$ & $14 \times 4$ & $120 \times 4$ & 60  $\times$ 12 & 3,7,8 \\
IRAS 10339+1548   & 2007 Jun 08 & BPU-offset & $14 \times 7$ & $14 \times 7$ & $30 \times 3$ & $30 \times 3$ & $120 \times 2$ & 60  $\times$ 2  & 1     \\
IRAS 10378+1109   &             & BPU-offset & $60 \times 2$ & $60 \times 2$ & $30 \times 3$ & $30 \times 3$ & $120 \times 3$ & 60  $\times$ 4  & 4     \\
IRAS 10485$-$1447 &             & BPU        & $60 \times 2$ & $60 \times 2$ & $30 \times 4$ & $30 \times 4$ & $120 \times 2$ & 240 $\times$ 2  & 5,6   \\
IRAS 11028+3130   & 2007 Jun 09 & BPU-offset & $60 \times 2$ & $60 \times 2$ & $30 \times 3$ & $30 \times 3$ & $120 \times 2$ & 60  $\times$ 2  & 1     \\
IRAS 11180+1623   & 2007 Jun 08 & BPU-offset & $14 \times 7$ & $14 \times 7$ & $30 \times 3$ & $30 \times 3$ & $120 \times 2$ & 60  $\times$ 2  & 1     \\
IRAS 11524+1058   & 2007 Jun 12 & BPU-offset & $14 \times 7$ & $14 \times 7$ & $30 \times 3$ & $30 \times 3$ & $120 \times 2$ & 60  $\times$ 2  & 1     \\
IRAS 12018+1941   &             & BPU-offset & $60 \times 1$ & $60 \times 1$ & $30 \times 3$ & $30 \times 3$ & $120 \times 3$ & 60  $\times$ 4  & 4     \\
IRAS 12032+1707   &             & BPU-offset & $60 \times 2$ & $60 \times 2$ & $30 \times 2$ & $30 \times 2$ & $120 \times 3$ & 240 $\times$ 3  & 4,6   \\
IRAS 12112+0305   &             & BPU-offset & $14 \times 3$ & $14 \times 3$ & $30 \times 2$ & $30 \times 2$ & $120 \times 2$ & 60  $\times$ 4  & 4     \\
IRAS 12540+5708   &             & BPU        & $14 \times 2$ & $14 \times 2$ & $6  \times 5$ & $6  \times 5$ & $30  \times 6$ & 60  $\times$ 4  & 9     \\
IRAS 13218+0552   &             & BPU-offset & $60 \times 1$ & $60 \times 1$ & $30 \times 3$ & $30 \times 3$ & $120 \times 3$ & 60  $\times$ 4  & 4     \\
IRAS 13428+5608   &             & BPU        & $14 \times 2$ & $14 \times 2$ & $14 \times 2$ & $14 \times 2$ & $30  \times 6$ & 60  $\times$ 4  & 4     \\
IRAS 13451+1232   &             & BPU        & $14 \times 3$ & $14 \times 3$ & $30 \times 2$ & $30 \times 2$ & $30  \times 6$ & 60  $\times$ 4  & 4     \\
IRAS 14059+2000   & 2007 Jul 31 & BPU-offset & $60 \times 2$ & $60 \times 2$ & $30 \times 3$ & $30 \times 3$ & $120 \times 2$ & 60  $\times$ 2  & 1     \\
IRAS 14070+0525   &             & BPU-offset & $60 \times 2$ & $60 \times 2$ & $30 \times 2$ & $30 \times 2$ & $120 \times 3$ & 240 $\times$ 2  & 4     \\
IRAS 14553+1245   & 2007 Jul 31 & BPU        & $60 \times 2$ & $60 \times 2$ & $30 \times 3$ & $30 \times 3$ & $120 \times 2$ & 60  $\times$ 2  & 1     \\
IRAS 15327+2340   &             & BPU        & $14 \times 3$ & $14 \times 3$ & $6  \times 5$ & $6  \times 5$ & $30  \times 6$ & 60  $\times$ 4  & 10     \\
IRAS 16090$-$0139 &             & BPU-offset & $60 \times 1$ & $60 \times 1$ & $30 \times 3$ & $30 \times 3$ & $120 \times 2$ & 60  $\times$ 4  & 4     \\
IRAS 16255+2801   & 2006 Sep 17 & BPU        & $60 \times 2$ & $60 \times 2$ & $30 \times 3$ & $30 \times 3$ & $120 \times 2$ & 60  $\times$ 2  & 1     \\
IRAS 16300+1558   &             & BPU-offset & $60 \times 2$ & $60 \times 2$ & $30 \times 5$ & $30 \times 5$ & $120 \times 4$ & 240 $\times$ 4  & 4,6   \\
IRAS 17207$-$0014 &             & BPU        & $14 \times 3$ & $14 \times 3$ & $14 \times 3$ & $30 \times 2$ & $30  \times 6$ & 60  $\times$ 4  & 4     \\
IRAS 18368+3549   & 2007 May 01 & BPU        & $60 \times 2$ & $60 \times 2$ & $30 \times 3$ & $30 \times 3$ & $120 \times 2$ & 60  $\times$ 2  & 1     \\
IRAS 18588+3517   & 2006 Nov 20 & BPU        & $60 \times 2$ & $60 \times 2$ & $30 \times 3$ & $30 \times 3$ & $120 \times 2$ & 60  $\times$ 2  & 1     \\
IRAS 20100$-$4156 &             & BPU-offset & $60 \times 1$ & $60 \times 1$ & $30 \times 2$ & $30 \times 2$ & $120 \times 2$ & 60  $\times$ 4  & 4     \\
IRAS 20286+1846   & 2006 Nov 20 & BPU        & $60 \times 2$ & $60 \times 2$ & $30 \times 3$ & $30 \times 3$ & $120 \times 2$ & 60  $\times$ 2  & 1     \\
IRAS 21077+3358   & 2007 Jun 13 & BPU        & $60 \times 2$ & $60 \times 2$ & $30 \times 3$ & $30 \times 3$ & $120 \times 2$ & 60  $\times$ 2  & 1     \\
IRAS 21272+2514   &             & BPU-offset & $60 \times 2$ & $60 \times 2$ & $30 \times 3$ & $30 \times 3$ & $120 \times 2$ & 240 $\times$ 1  & 4,11  \\
IRAS 22055+3024   & 2007 Jun 27 & BPU        & $60 \times 2$ & $60 \times 2$ & $30 \times 3$ & $30 \times 3$ & $120 \times 2$ & 60  $\times$ 2  & 1     \\
IRAS 22116+0437   & 2006 Dec 21 & BPU-offset & $60 \times 2$ & $60 \times 2$ & $30 \times 3$ & $30 \times 3$ & $120 \times 2$ & 60  $\times$ 2  & 1     \\
IRAS 22491$-$1808 &             & BPU-offset & $60 \times 1$ & $60 \times 1$ & $30 \times 2$ & $30 \times 2$ & $120 \times 2$ & 60  $\times$ 4  & 4     \\
IRAS 23028+0725   & 2006 Dec 20 & BPU        & $14 \times 7$ & $14 \times 7$ & $30 \times 3$ & $30 \times 3$ & $120 \times 2$ & 60  $\times$ 2  & 1     \\
IRAS 23233+0946   &             & BPU        & $60 \times 2$ & $60 \times 2$ & $30 \times 4$ & $30 \times 4$ & $120 \times 5$ & 240 $\times$ 4  & 5,6   \\
IRAS 23365+3604   &             & BPU        & $14 \times 3$ & $14 \times 3$ & $30 \times 2$ & $30 \times 2$ & $30  \times 6$ & 60  $\times$ 4  & 4     \\
\hline
IRAS 00163$-$1039 &             & BPU        & $14 \times 3$ & $14 \times 3$ & $14 \times 2$ & $14 \times 2$ & $30  \times 3$ & 60  $\times$ 2  & 3     \\
IRAS 01572+0009   &             & BPU        & $14 \times 3$ & $14 \times 3$ & $30 \times 2$ & $30 \times 2$ & $30  \times 6$ & 60  $\times$ 4  & 4     \\
IRAS 05083+7936   &             & BPU        & $14 \times 6$ & $14 \times 6$ & $14 \times 4$ & $14 \times 4$ & $30  \times 6$ & 60  $\times$ 2  & 3     \\
IRAS 06538+4628   &             & BPU        & $14 \times 3$ & $14 \times 3$ & $14 \times 2$ & $14 \times 2$ & $30  \times 3$ & 60  $\times$ 2  & 3     \\
IRAS 08559+1053   &             & BPU-offset & $60 \times 2$ & $60 \times 2$ & $30 \times 3$ & $30 \times 3$ & $120 \times 3$ & 60  $\times$ 4  & 12    \\
IRAS 09437+0317   &             & BPU        & $60 \times 2$ & $60 \times 2$ & $30 \times 3$ & $30 \times 3$ & $120 \times 3$ & 240 $\times$ 2  & 3     \\
IRAS 10565+2448   &             & BPU        & $14 \times 3$ & $14 \times 3$ & $30 \times 2$ & $30 \times 2$ & $30  \times 6$ & 60  $\times$ 4  & 4     \\
IRAS 11119+3257   &             & BPU        & $60 \times 1$ & $60 \times 1$ & $30 \times 3$ & $30 \times 3$ & $120 \times 3$ & 60  $\times$ 4  & 4     \\
IRAS 13349+2438   &             & BPU        & $14 \times 5$ & $14 \times 5$ & $14 \times 5$ & $14 \times 5$ & $120 \times 5$ & 60  $\times$ 10 & 13    \\
IRAS 15001+1433   &             & BPU-offset & $60 \times 2$ & $60 \times 2$ & $30 \times 3$ & $30 \times 3$ & $120 \times 3$ & 60  $\times$ 4  & 4     \\
IRAS 15206+3342   &             & BPU-offset & $60 \times 1$ & $60 \times 1$ & $30 \times 3$ & $30 \times 3$ & $120 \times 3$ & 60  $\times$ 4  & 4     \\
IRAS 20460+1925   &             & BPU        & $14 \times 5$ & $14 \times 5$ & $14 \times 5$ & $14 \times 5$ & $120 \times 5$ & 60  $\times$ 10 & 13    \\
IRAS 23007+0836   &             & BPU        & $14 \times 2$ & $14 \times 2$ & $6  \times 5$ & $6  \times 5$ & $30  \times 4$ & 60  $\times$ 2  & 14    \\
IRAS 23394$-$0353 &             & BPU        & $30 \times 6$ & $30 \times 6$ & $60 \times 2$ & $60 \times 2$ & $14  \times 6$ & 14  $\times$ 4  & 3     \\
IRAS 23498+2423   &             & BPU-offset & $60 \times 2$ & $60 \times 2$ & $30 \times 2$ & $30 \times 2$ & $120 \times 3$ & 240 $\times$ 2  & 4     \\
\enddata
\tablecomments{\textit{Spitzer} archival data are from programs: (1) 30407 (PI: J.~Darling); (2) - 3237 (PI: E.~Sturm); (3) - 30323 (PI: L.~Armus); (4) - 105 (PI: J.~Houck); (5) - 2306 (PI: M.~Imanishi); (6) - 3187 (PI: S.~Veilleux); (7) - 3605 (PI: C.~Bradford); (8) - 20549 (PI: R.~Joseph); (9) - 1442 (PI: L.~Armus); (10) - 1444 (PI: L.~Armus); (11) - 20375 (PI: L.~Armus); (12) - 666 (PI: J.~Houck); (13) - 61 (PI: G.~Rieke); (14) - 14 (PI: J.~Houck). Exposure times for all modules are given as seconds per cycle $\times$ number of cycles.}
\end{deluxetable*}


\section{Data reduction}

\subsection{Dedicated observations of OHM galaxies}

The data were processed using the \textit{Spitzer} Science Center S17.0 data pipeline. We used basic calibrated data (BCD) products for our analysis, having already been corrected for flat-fielding, stray light contributions, non-linear responsivity in the pixels, and ``drooping'' (an increase in detector pixel voltage that occurs during non-destructive readouts). The 2-D BCD images were first medianed over the data cycles at each nod position to remove transient effects such as cosmic rays. For the SL and LL modules, we subtracted the sky contribution by differencing the BCD images for each nod position with the adjacent position in the same module. 

The slit sizes of the SH and LH modules are too small to permit extraction of a sky background during the same observation. The continuum levels in the high-resolution modules, however, contain strong contributions from scattered zodiacal light. Estimations of the flux at 15~\um~using SPOT predict zodi contributions within the {\it Spitzer} beam ranging from 20--80~mJy, which in many cases is of comparable magnitude to the expected signal from the galaxies themselves. The wavelength-dependent brightness of the sky contribution means that it cannot be corrected using a simple scaling, and so we did not attempt to further calibrate either HR module. The calibrated low-resolution spectra were thus used for absolute fluxes and the high-resolution spectra for line ratio diagnostics (which are unaffected by continuum levels). 

To obtain more accurate measurements of faint lines at longer wavelengths, we took dedicated off-source sky observations in the LH module for all 24 OHMs in our program. Subtraction of the wavelength-dependent background, however, significantly affected the measured line fluxes. For galaxies with background subtraction in only the LH module, this changes the line ratios measured between different HR modules ({\it eg}, \sIII$\lambda33$/\sIV$\lambda10.5$). We reduced the data both with and without subtraction of the sky backgrounds; the background-subtracted LH spectra are attached as an appendix. 

Following the initial cleaning of the 2-D BCD products, we eliminated rogue pixels using the IDL package IRSCLEAN\_MASK\footnote[1]{IRSCLEAN\_MASK is available at \texttt{http://ssc.spitzer.caltech.edu/archanaly/\\contributed/irsclean/IRSCLEAN\_MASK.html}}. We used rogue pixel masks provided by the SSC for each IRS campaign, and supplemented the standard masks with manual cleaning of each nod and module. The 1-D spectra were then extracted using the Spitzer IRS Custom Extractor (SPICE) v.2.0. For all modules we used the optimal extraction routine with the standard aperture to improve the S/N ratio in faint galaxies. 

The low-resolution modules were stitched together to match continuum levels by using a multiplicative scaling. We fixed the LL1 module and then scaled LL2 to LL1, SL1 to LL2, and SL2 to SL1. The mean scaling factors were $1.03\pm0.08$ for LL2 to LL1, $1.49\pm0.69$ for SL1 to LL2, and $1.37\pm0.78$ for SL2 to SL1. We then calibrated the entire low-resolution spectra as a single unit by scaling to the IRS 22~\um~sample-up-the-ramp (SUR) peakups. The required scaling in the majority of cases was quite small, indicating that the sky subtraction and spectral extraction techniques are robust; the mean scaling factor was $0.94\pm0.08$. The accuracy of the overall continuum flux calibration is $\sim5\%$ \citep{hou04}. 

Noisy areas on both ends of the SH and LH orders were trimmed from the 1-D spectra. These areas typically encompass a range of 10--30 pixels on the edges of the orders and correspond to areas of decreased sensitivity on the detector. We deliberately trimmed only pixels with an overlapping wavelength range in adjacent orders so that a maximum amount of information is preserved. In isolated cases, we also removed obvious rogue pixels by hand from spectra in which exceptional 1-channel features appear in only a single nod. 

We calculated a simple figure of merit to measure the signal-to-noise ratio in the low-resolution data. The data near $\lambda_{rest}=21$~\um~(a feature-free area near the center of most spectra) are fit with a low-order polynomial; the median flux in that region is then divided by the rms noise to yield the S/N (Table~\ref{tbl-puflux}). We note that this parameter is a function of wavelength, as well as the performance and integration time in each spectral module; this is intended to give only a rough estimate for each object. The S/N for the samples ranges from $\sim10-110$, with a median of 35. 

\subsection{Archival data}

Since the archival OHM and non-masing galaxies did not come from a unified observing program, the version of the \textit{Spitzer} data pipeline and the level of processing varied slightly from object to object - we used the most recent versions available in the archive (v15.3.0 or later). The reduction process was identical to that for the OHM galaxies in our program, with the exception of the LR photometric scaling; since observations in the archive varied in availability of peakup data, we used a variety of sources to calibrate the spectra. In order of priority, we used the IRS dedicated 22~\um~SUR peakups, IRS acquisition 16 and 22~\um~DCS peakups, or IRAS 25~\um~observations (all lying within the coverage of the LL modules). For galaxies from the archive, 17 are calibrated with DCS peakups, 21 with IRAS 25~\um~photometry, and 4 are left uncalibrated. The mean scaling factor for the objects without SUR photometry was $1.05\pm0.25$, slightly higher than the mean scaling for galaxies in the dedicated OHM program. 

Five OHMs and five non-masing galaxies from the archive had both SH and LH sky backgrounds taken simultaneously with the spectroscopic observations; the remainder had no high-resolution sky backgrounds in either module. Since we emphasize uniformity of the observations to the fullest extent possible, all data used for statistical comparisons between the samples use data \emph{without} HR sky subtraction; line measurements for the background-subtracted objects are given in the Appendix. 

The spectra for IRAS~20460+1925 and IRAS~23028+0725 had no flux in the SL modules, most likely due to a pointing error during observations. No SL data from either galaxy are used in our analyses here or in Paper~II. 


\section{Results}\label{sec-results}

\begin{figure*}
\includegraphics[scale=0.8]{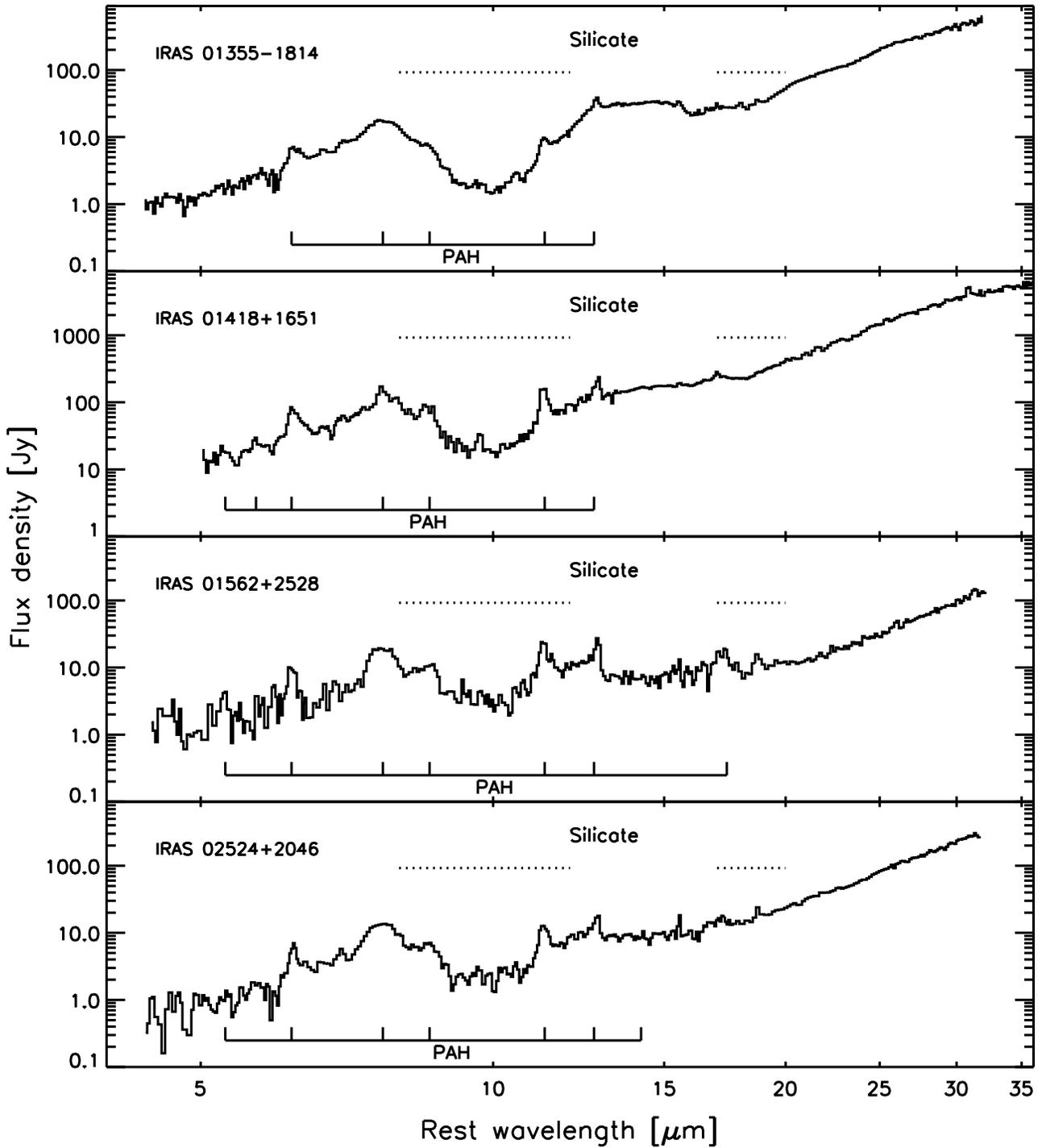}
\caption{IRS spectra from the low-resolution modules (LR) for OHMs. Spectra for all OHMs and non-masing galaxies are available as an online supplement; portions are shown here for guidance on form and style. All detected PAH emission and absorption features from water ice and silicates are marked. \label{fig-lr1}}
\end{figure*}

\begin{figure*}
\includegraphics[scale=0.8]{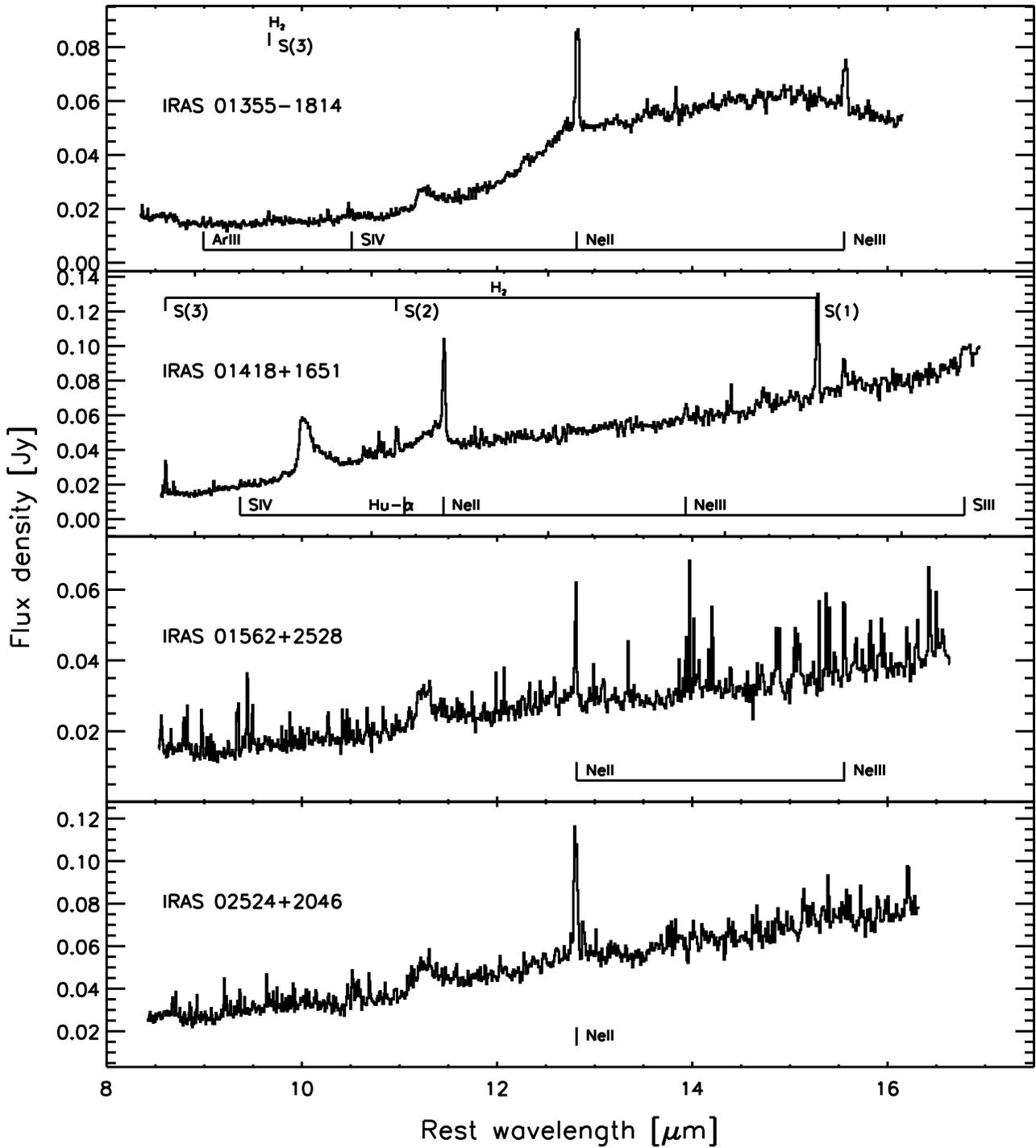}
\caption{IRS spectra from the short-high module (SH) for OHMs. Spectra for all OHMs and non-masing galaxies are available as an online supplement; portions are shown here for guidance on form and style. All detected atomic and \htwo~features in each spectra are marked. \label{fig-sh1}}
\end{figure*}

\begin{figure*}
\includegraphics[scale=0.8]{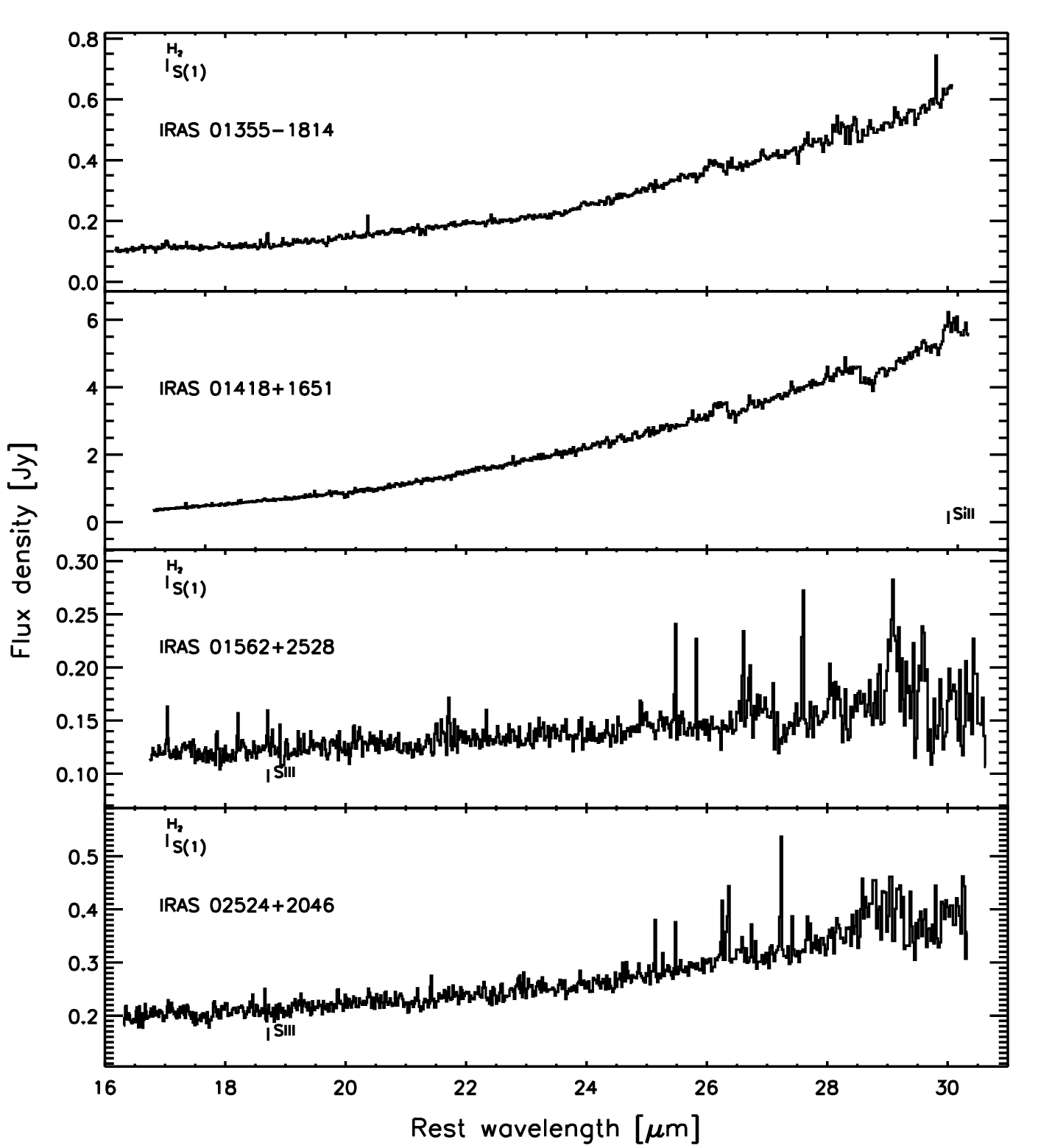}
\caption{IRS spectra from the long-high module (LH) for OHMs. Spectra for all OHMs and non-masing galaxies are available as an online supplement; portions are shown here for guidance on form and style. All detected atomic and \htwo~emission features are marked. \label{fig-lh1}}
\end{figure*}

We show examples of the peakup-scaled low-resolution spectra for the OHMs in Figure~\ref{fig-lr1}, with the individual modules stitched together and bonus orders removed. Examples of the high-resolution SH and LH data are shown in Figures~\ref{fig-sh1} and \ref{fig-lh1}; full spectra for all galaxies are available as an online supplement. While individual orders within the high-resolution modules are typically well-aligned in flux, the differences in calibration between the SH and LH modules are clearly apparent when matching the spectra; this is due to a combination of different slit sizes for the SH and LH modules (a factor of $\sim4$) and the lack of separate sky subtraction for the SH modules. For this reason, as well as emphasizing the narrow atomic and molecular features visible in the high-resolution spectra, we display separate plots for the SH and LH modules. 

A small amount of the archival objects have previously published full IRS spectra \citep{arm04,wee05,arm07}, mainly consisting of bright, nearby galaxies. \citet{far07} published HR spectra for roughly half of our archival OHM hosts. While many papers use the available data from the GTO programs, however, the majority of objects extracted from the archive have {\em no} published spectra, although some data are used in larger studies of ULIRG properties \citep[{\it eg},][]{hig06,des07,hao07}. The spectra for many of the archival galaxies are thus presented here for the first time. 

Comparison of our data with the few spectra of objects previously published ({\it eg}, Mrk~1014, Arp~220, NGC~7469) revealed no significant differences in spectral shape or detection of individual features. Measurements of line flux and equivalent widths, however, may be affected by the photometric scaling and/or line-fitting routines used; for this reason, we chose to reduce {\em all} data in a uniform matter.

For many of our high-resolution spectra, especially those with low S/N, there exist individual spikes that do not correspond to any identified feature (see IRAS~01562+2528 for a prominent example). These features are typically 1--2 channels wide, much narrower than the expected linewidth for an unresolved feature. We regard such features as spurious, possibly caused by hot pixels or other instrumental conditions that are not corrected by our cleaning routines. All features we regard as valid detections are listed in the data tables, with the locations of the most common features marked on the spectra themselves.  

\subsection{Continuum} \label{ssec-cont}

\begin{figure}
\includegraphics[scale=0.5]{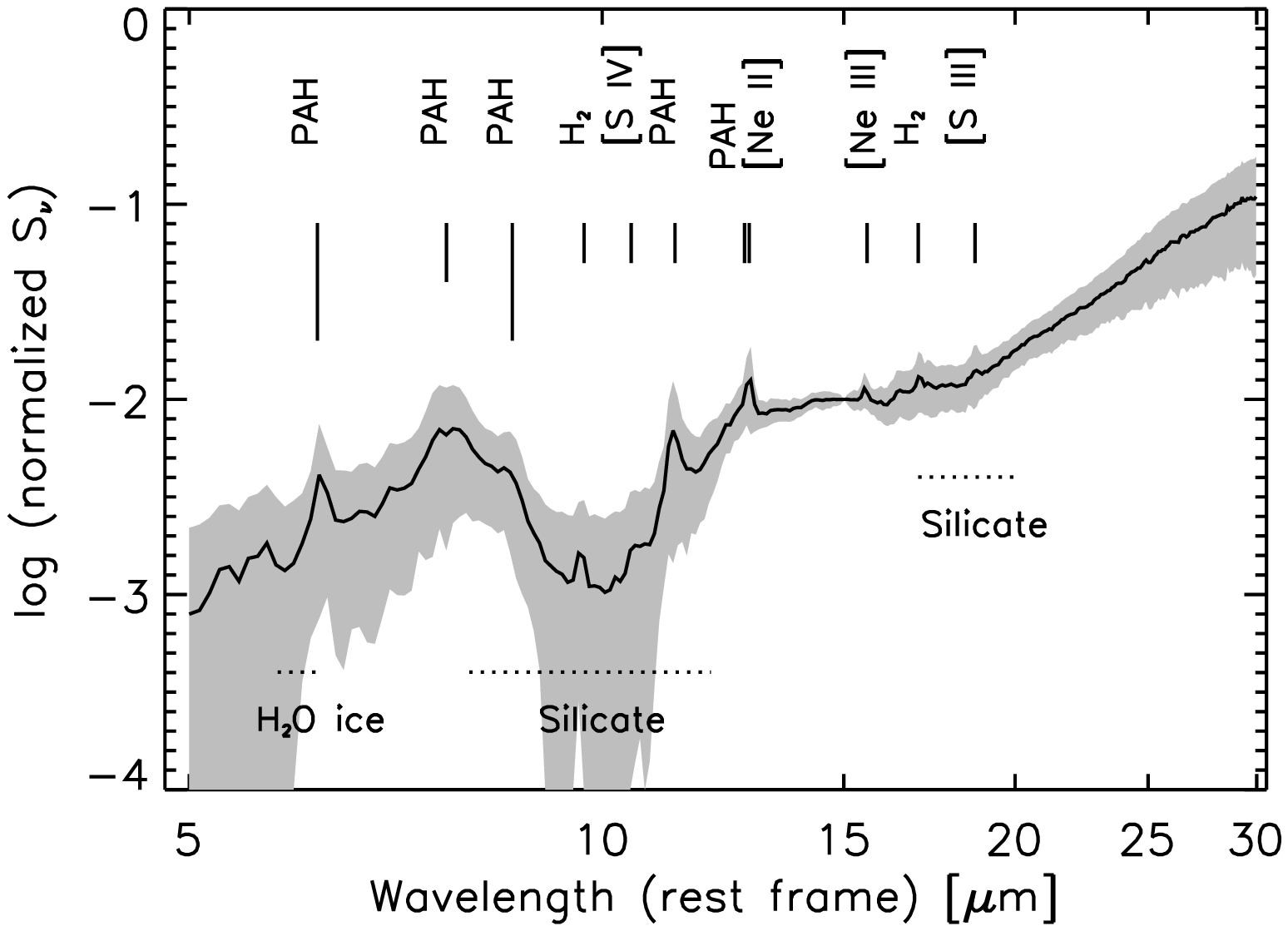}
\caption{Low-resolution spectrum of all galaxies in our sample normalized at $S_\lambda=15~\mu$m. The black spectrum is the composite template made from an error-weighted median of the individual galaxies; the grey shaded area shows the 1-$\sigma$ envelope for each resolution element (exaggerated toward negative values in log space). \label{fig-lravg}}
\end{figure}

The continuum emission for all objects in the OHM sample has a relatively homogenous spectral shape over the range of the IRS, although differences in spectral shape between the two samples do appear and are explored in Paper II. Figure~\ref{fig-lravg} shows the individual objects overlaid with a template generated by medianing the flux in each wavelength bin from \emph{all} galaxies. The template bears a close resemblance to starbursting ULIRG spectra seen in previous surveys \citep{hao07,wee09}. The LR spectrum clearly shows silicate absorption at 9.7 and 18~\um~and water ice absorption at 6~\um. Low-resolution emission features are dominated by the broad PAH features from 6--13~\um, with weaker contibutions from neon, sulfur, and molecular hydrogen also visible. 

The continuum data from $\lambda_{rest}=20-30$~\um~is in most cases well-characterized by a power-law fit, with the short wavelength break occurring near the 18~\um~silicate feature and the long wavelength end cut off by the spectral range of the IRS. Shortward of 15~\um, the continuum becomes increasingly contaminated by individual absorption and emission features, especially from PAH emission and the deep silicate absorption at 9.7~\um. Following \citet{bra06}, we fit a spectral index to the continuum in two components, with $\alpha_{30-20}$ measuring the relatively feature-free flux from $20-30$~\um~and $\alpha_{15-6}$ measuring the contribution from $5.3-14.8$~\um; the wavelengths are slightly shifted to avoid contamination from water ice at 6~\um~and [\ion{Ne}{3}] at 15.6~\um~(Table~\ref{tbl-puflux}). 

\begin{deluxetable*}{lrrcrrr}
\tabletypesize{\scriptsize}
\tablecaption{IRS photometry and continuum measurements \label{tbl-puflux}}
\tablewidth{0pt}
\tablehead{
\colhead{Object} & 
\colhead{16~\um~PU} &
\colhead{22~\um~PU} & 
\colhead{PU type} &
\colhead{$\alpha_{15-6}$} &
\colhead{$\alpha_{30-20}$} &
\colhead{S/N}
\\
\colhead{} & 
\colhead{[mJy]} & 
\colhead{[mJy]} & 
\colhead{} & 
\colhead{} &
\colhead{} &
\colhead{} 
}
\startdata
IRAS 01355$-$1814 & \ldots & 45.5   & DCS     & $2.8\pm0.3$  & $5.2\pm0.8$   &  75 \\
IRAS 01418+1651   & \ldots & \ldots & \ldots  & $2.2\pm0.3$  & $5.2\pm0.8$   &  18 \\
IRAS 01562+2528   & 6.1    & 13.0   & SUR     & $1.8\pm0.3$  & $5.1\pm0.8$   &  11 \\
IRAS 02524+2046   & 10.6   & 21.7   & SUR     & $2.0\pm0.4$  & $5.7\pm1.0$   &  25 \\
IRAS 03521+0028   & 25.6   & \ldots & DCS     & $2.3\pm0.4$  & $5.5\pm0.8$   &  63 \\
IRAS 04121+0223   & 10.5   & 22.6   & SUR     & $2.6\pm0.1$  & $5.9\pm0.5$   &  21 \\
IRAS 04454$-$4838 & \ldots & \ldots & \ldots  & $2.7\pm0.4$  & $6.1\pm0.9$   &  37 \\
IRAS 06487+2208   & 85.0   & 177.5  & SUR     & $2.4\pm0.2$  & $3.6\pm0.8$   &  48 \\
IRAS 07163+0817   & 12.3   & 28.8   & SUR     & $2.5\pm0.3$  & $4.8\pm1.0$   &  26 \\
IRAS 07572+0533   & 52.1   & 103.0  & SUR     & $2.6\pm0.3$  & $3.0\pm0.7$   &  55 \\
IRAS 08201+2801   & 61.8   & 74.3   & SUR     & $2.2\pm0.2$  & $4.5\pm0.7$   &  37 \\
IRAS 08449+2332   & 26.6   & 48.2   & SUR     & $2.2\pm0.2$  & $4.4\pm0.4$   &  52 \\
IRAS 08474+1813   & 9.5    & 28.7   & SUR     & $2.2\pm0.5$  & $6.5\pm0.6$   &  30 \\
IRAS 09039+0503   & 86.7   & \ldots & DCS     & $1.8\pm0.5$  & $5.5\pm0.7$   &  41 \\
IRAS 09539+0857   & \ldots & \ldots & \ldots  & $2.2\pm0.1$  & $6.0\pm0.3$   &  36 \\
IRAS 10035+2740   & 9.2    & 22.8   & SUR     & $2.1\pm0.2$  & $6.0\pm0.5$   &  23 \\
IRAS 10039$-$3338 & \ldots & \ldots & \ldots  & $-0.1\pm0.1$ & $5.4\pm0.3$   &  17 \\
IRAS 10173+0828   & \ldots & \ldots & \ldots  & $2.6\pm0.4$  & $6.4\pm1.0$   &  48 \\
IRAS 10339+1548   & 10.7   & 28.1   & SUR     & $2.5\pm0.3$  & $5.1\pm0.7$   &  22 \\
IRAS 10378+1109   & \ldots & 114.8  & DCS     & $2.0\pm0.3$  & $4.5\pm0.7$   &  35 \\
IRAS 10485$-$1447 & \ldots & \ldots & \ldots  & $2.5\pm0.3$  & $4.8\pm0.8$   &  75 \\
IRAS 11028+3130   & 4.8    & 13.0   & SUR     & $2.4\pm0.2$  & $7.0\pm0.6$   &  26 \\
IRAS 11180+1623   & 14.8   & 33.6   & SUR     & $2.2\pm0.3$  & $5.6\pm0.9$   &  15 \\
IRAS 11524+1058   & 8.2    & 14.8   & SUR     & $1.5\pm0.4$  & $6.1\pm0.8$   &  20 \\
IRAS 12018+1941   & 121.7  & \ldots & DCS     & $3.0\pm0.3$  & $2.9\pm0.4$   &  63 \\
IRAS 12032+1707   & \ldots & 73.9   & DCS     & $2.2\pm0.3$  & $4.2\pm0.7$   &  42 \\
IRAS 12112+0305   & 91.9   & \ldots & DCS     & $2.3\pm0.3$  & $5.3\pm0.7$   &  42 \\
IRAS 12540+5708   & \ldots & \ldots & \ldots  & $1.6\pm0.3$  & $2.4\pm0.7$   &  30 \\
IRAS 13218+0552   & 212.9  & \ldots & DCS     & $0.6\pm0.3$  & $2.4\pm0.7$   & 100 \\
IRAS 13428+5608   & \ldots & \ldots & \ldots  & $1.9\pm1.3$  & $4.6\pm0.6$   &  12 \\
IRAS 13451+1232   & \ldots & \ldots & \ldots  & $2.3\pm0.4$  & $1.9\pm0.7$   &  50 \\
IRAS 14059+2000   & 12.2   & 26.3   & SUR     & $1.1\pm0.6$  & $4.9\pm0.8$   &  20 \\
IRAS 14070+0525   & \ldots & 30.1   & DCS     & $1.6\pm0.3$  & $5.6\pm0.5$   &  33 \\
IRAS 14553+1245   & 28.7   & 61.2   & SUR     & $2.0\pm0.7$  & $4.3\pm0.7$   &  22 \\
IRAS 15327+2340   & \ldots & \ldots & \ldots  & $2.4\pm0.2$  & $5.5\pm0.4$   &  17 \\
IRAS 16090$-$0139 & 71.4   & \ldots & DCS     & $1.3\pm0.4$  & $4.6\pm0.6$   &  59 \\
IRAS 16255+2801   & 16.0   & 36.7   & SUR     & $1.4\pm0.3$  & $4.7\pm0.6$   &  22 \\
IRAS 16300+1558   & \ldots & 35.0   & DCS     & $1.7\pm0.5$  & $5.6\pm0.9$   &  31 \\
IRAS 17207$-$0014 & \ldots & \ldots & \ldots  & $1.8\pm0.6$  & $6.2\pm0.6$   &  28 \\
IRAS 18368+3549   & 26.6   & 49.8   & SUR     & $1.7\pm0.5$  & $5.4\pm0.8$   &  29 \\
IRAS 18588+3517   & 43.6   & 92.1   & SUR     & $1.5\pm0.3$  & $4.8\pm0.8$   &  28 \\
IRAS 20100$-$4156 & 86.7   & \ldots & DCS     & $1.9\pm0.2$  & $5.2\pm0.7$   &  73 \\
IRAS 20286+1846   & 11.4   & 22.0   & SUR     & $2.3\pm0.3$  & $5.8\pm0.6$   &  20 \\
IRAS 21077+3358   & 31.1   & 62.1   & SUR     & $2.8\pm0.3$  & $4.2\pm0.6$   &  52 \\
IRAS 21272+2514   & \ldots & 39.0   & DCS     & $2.0\pm0.3$  & $5.0\pm0.7$   &  36 \\
IRAS 22055+3024   & 54.5   & 132.4  & SUR     & $2.7\pm0.3$  & $4.0\pm0.7$   &  30 \\
IRAS 22116+0437   & 46.6   & 68.2   & SUR     & $2.3\pm0.5$  & $4.5\pm0.9$   &  35 \\
IRAS 22491$-$1808 & 87.5   & \ldots & DCS     & $2.8\pm0.3$  & $5.1\pm0.6$   &  28 \\
IRAS 23028+0725   & 56.4   & 140.7  & SUR     & \ldots       & $3.2\pm0.7$   &  25 \\
IRAS 23233+0946   & \ldots & \ldots & \ldots  & $2.1\pm0.3$  & $4.7\pm0.5$   &  33 \\
IRAS 23365+3604   & \ldots & \ldots & \ldots  & $2.5\pm3.1$  & $4.6\pm0.5$   &  15 \\
\hline                                        
IRAS 00163$-$1039 & \ldots & \ldots & \ldots  & $2.3\pm0.4$ & $2.1\pm0.4$   &  34 \\
IRAS 01572+0009   & \ldots & \ldots & \ldots  & $1.8\pm0.2$ & $2.2\pm0.3$   &  81 \\
IRAS 05083+7936   & \ldots & \ldots & \ldots  & $2.0\pm0.4$ & $2.8\pm0.4$   &  48 \\
IRAS 06538+4628   & \ldots & \ldots & \ldots  & $2.9\pm0.3$ & $2.5\pm0.4$   &  37 \\
IRAS 08559+1053   & \ldots & 90.7   & DCS     & $1.3\pm0.1$ & $2.8\pm0.4$   &  64 \\
IRAS 09437+0317   & \ldots & \ldots & \ldots  & $1.7\pm0.3$ & $2.2\pm0.4$   &  19 \\
IRAS 10565+2448   & \ldots & \ldots & \ldots  & $2.1\pm0.4$ & $3.2\pm0.5$   &  19 \\
IRAS 11119+3257   & \ldots & \ldots & \ldots  & $1.2\pm0.1$ & $2.5\pm0.3$   &  79 \\
IRAS 13349+2438   & \ldots & \ldots & \ldots  & $0.8\pm0.1$ & $0.1\pm0.1$   & 110 \\
IRAS 15001+1433   & \ldots & 135.3  & DCS     & $1.9\pm0.2$ & $3.4\pm0.5$   &  58 \\
IRAS 15206+3342   & 110.7  & \ldots & DCS     & $2.3\pm0.4$ & $2.6\pm0.4$   &  65 \\
IRAS 20460+1925   & \ldots & \ldots & \ldots  &  \ldots     & $1.1\pm0.3$   &  16 \\
IRAS 23007+0836   & \ldots & \ldots & \ldots  & $1.7\pm0.1$ & $1.8\pm0.4$   &  19 \\
IRAS 23394$-$0353 & \ldots & \ldots & \ldots  & $1.8\pm0.3$ & $2.9\pm0.4$   &  37 \\
IRAS 23498+2423   & 46.5   & \ldots & DCS     & $1.0\pm0.1$ & $3.1\pm0.4$   &  87 \\
\enddata
\tablecomments{Errors in the peakup (PU) fluxes are at the 15\% level.} 
\end{deluxetable*}

The mean 15-6 spectral index for the entire OHM sample is $\alpha_{15-6}=2.1\pm0.6$; the mean 30-20 slope is $\alpha_{30-20}=4.8\pm1.1$. The shallowest 15-6 slope occurs for IRAS~10039$-$3338 ($\alpha_{15-6}=-0.1$), an object with weak PAH features and very strong silicate absorption; the steepest index occurs for IRAS~12018+1941 ($\alpha_{15-6}=3.0$), which has moderate PAH and line emission features and a nearly constant spectral index over the entire mid-IR range. Steeper 15-6 indices are likely due to a combination of smaller relative quantities of warm dust (thermal blackbodies of $\sim300$~K peaking near 10~\um) and larger quantities of cooler dust. 

The shallowest 30-20 slope occurs for IRAS~13451+1232 ($\alpha_{30-20}=1.9$); the low-resolution spectrum for this object more closely resembles that seen in Seyfert galaxies and PG quasars \citep{sch06,hao07}, with weak PAH emission and shallower silicate absorption. The continuum emission for this object is also much closer to being uniform over the entire range of the IRS; the difference between the two spectral indices is only $\Delta\alpha=-0.4$, compared to an average of $\Delta\alpha=2.7$ for the entire sample. This behavior is more typical of non-thermal emission that can extend over many decades with the same index. The steepest 30-20 emission measured is from IRAS~11028+3130 ($\alpha_{30-20}=7.0$); the galaxy shows moderate PAH and line emission features, but a flat continuum between 12 and 20~\um. 

The non-masing galaxies have average spectral indices of $\alpha_{15-6}=1.8\pm0.6$ and $\alpha_{30-20}=2.5\pm0.9$. A full statistical analysis of the values for the two samples (particularly $\alpha_{30-20}$) is presented in Paper~II. 

\subsection{Atomic emission lines}\label{ssec-atomic}

We measured emission from atomic and molecular lines using the standard packages in the Spectroscopic Modeling Analysis and Reduction Tool (SMART) v6.2.4 \citep{hig04}. A simple Gaussian is a good fit for virtually all high-resolution lines in the sample; in cases where lines are blended (such as the \neV/\clII~and \oIV/\feII~complexes), we used a multi-Gaussian fit centered at the redshifted rest wavelengths of the expected transitions. To compute upper limits for non-detections, we use the 3-$\sigma$ noise measured from the surrounding continuum and a Gaussian shape with an FWHM estimated from detected lines. Accuracy for all measured line fluxes is on the order of $\sim10\%$ (Tables~\ref{tbl-hrlines1} and \ref{tbl-hrlines2}). 

The most common lines detected are the forbidden \neII~$\lambda12.814$ and \neIII~$\lambda15.555$ transitions. \neII~is observed in nearly the entire sample, with detections in 50/51 OHMs and 14/15 non-masing galaxies. The only exceptions are the OHM IRAS~11028+3130 and the non-maser IRAS~20460+1925. \neIII~is also common, detected in 43/51 OHMs and 14/15 non-masing galaxies. Other common lines are the \sIII~$\lambda18.713$ (detected in $\sim80\%$ of galaxies) and \sIV~$\lambda10.511$ ($\sim50\%$). \arIII~$\lambda8.991$ is detected in 15 OHMs and 5 non-masing galaxies, but the redshifted line is not visible in the SH module for archived objects at $z<0.1$. 

We detect ``rarer'' line transitions that appear in less than 15\% of our sample, including \neV~$\lambda14.322$ and $\lambda24.318$, \feII~$\lambda17.936$ and $\lambda25.988$, and \oIV~$\lambda25.890$. IRAS~10339+1548 is the only galaxy in the sample with detections for all atomic transitions listed above. Eight galaxies also show the non-forbidden \HI~7-6 transition (Humphreys-$\alpha$) at 12.368~\um. 

\sIII$\lambda33.481$ and \siII$\lambda34.815$ are commonly observed in ULIRGs for which the redshifted transitions appear within the red edge of the LH module. \citet{far07} detected \sIII$\lambda33.481$~in roughly half of all ULIRGs with $z<0.06$, all of which show the stronger \sIII$\lambda18.713$ associated transition. For our sample, however, many galaxies have the lines redshifted beyond the range of the IRS; when visible, the lines lie at the far red edge of the 11$^\textrm{\scriptsize{th}}$ order in the LH module, an area with high noise and decreased sensitivity with respect to neighboring orders. 

For galaxies taken from the {\it Spitzer} archive, we compared our high-resolution measurements with those appearing in the sample of \citet{far07}. The agreement between detected lines is good for nearly all objects; however, \citet{far07} report detections of \neV~in IRAS~11119+3257, \oIV~in IRAS~13451+1232, and \htwo~S(0) and S(2) lines in IRAS~01572+0009 and IRAS~23498+2423 which we fail to confirm. Limits for the measured fluxes are given in Tables~\ref{tbl-hrlines1} and \ref{tbl-hrlines2} and \ref{tbl-h2}.

\begin{deluxetable*}{lrrrrrrr}
\tabletypesize{\scriptsize}
\tablecaption{Atomic line fluxes for high-resolution spectra: $8-16$~microns\label{tbl-hrlines1}}
\tablewidth{0pt}
\tablehead{
\colhead{Object} & 
\colhead{[Ar \tiny{III}\scriptsize]} &
\colhead{[S \tiny{IV}\scriptsize]} &
\colhead{H\tiny{I 7-6}} &
\colhead{[Ne \tiny{II}\scriptsize]} &
\colhead{[Ne \tiny{V}\scriptsize]} &
\colhead{[Cl \tiny{II}\scriptsize]} &
\colhead{[Ne \tiny{III}\scriptsize]}
\\
\colhead{$\lambda_{rest}$ [\um]} & 
\colhead{8.991} & 
\colhead{10.511} & 
\colhead{12.368} & 
\colhead{12.814} & 
\colhead{14.322} & 
\colhead{14.369} & 
\colhead{15.555}
}
\hline    
\startdata
IRAS 01355$-$1814 & 0.15    & 0.10    & $<0.20$ &   2.45  & $<0.27$ & $<0.25$ & 0.89    \\
IRAS 01418+1651   & $-$     & 0.15    & 0.13    &   4.17  & $<0.46$ & $<0.29$ & 0.41    \\
IRAS 01562+2528   & $<0.77$ & $<1.93$ & $<0.30$ &   1.07  & $<0.91$ & $<0.75$ & 0.77    \\
IRAS 02524+2046   & $<0.63$ & $<1.01$ & $<0.36$ &   2.39  & $<0.68$ & $<0.56$ & $<0.97$ \\
IRAS 03521+0028   & $<0.34$ & $<0.40$ & $<0.41$ &   2.72  & $<0.36$ & $<0.24$ & 1.11    \\
IRAS 04121+0223   & 0.35    & $<0.84$ & $<0.41$ &   1.81  & $<1.01$ & $<0.99$ & $<0.87$ \\
IRAS 04454$-$4838 & $-$     & 0.47    & $<0.74$ &   2.04  & $<0.75$ & $<0.61$ & 0.39    \\
IRAS 06487+2208   & 1.71    & 1.78    & $<0.81$ &  11.75  & $<1.44$ & $<1.19$ & 9.45    \\
IRAS 07163+0817   & 0.44    & $<1.05$ & $<0.22$ &   3.19  & $<0.46$ & $<0.39$ & 0.33    \\
IRAS 07572+0533   & $<1.69$ & $<0.87$ & $<0.40$ &   1.74  & $<0.79$ & $<0.65$ & $<0.89$ \\
IRAS 08201+2801   & $<0.70$ & $<2.74$ & $<0.64$ &   2.23  & $<0.60$ & $<0.41$ & 0.88    \\
IRAS 08449+2332   & $<0.94$ & $<0.71$ & $<0.64$ &   3.60  & $<2.03$ & $<1.64$ & 1.57    \\
IRAS 08474+1813   & $<0.76$ & $<0.64$ & $<0.36$ &   1.26  & $<1.10$ & $<0.89$ & $<1.10$ \\
IRAS 09039+0503   & 0.11    & $<0.27$ & $<0.85$ &   3.68  & $<0.46$ & $<0.38$ & 1.17    \\
IRAS 09539+0857   & $<0.63$ & $<0.43$ & $<0.30$ &   1.25  & $<0.86$ & $<0.69$ & $<1.88$ \\
IRAS 10035+2740   & $<0.89$ & $<0.75$ & $<0.48$ &   1.82  & $<0.42$ & $<0.31$ & 0.57    \\
IRAS 10039$-$3338 & $-$     & 0.93    & $<1.25$ &  16.69  & $<1.21$ & $<1.10$ & 3.93    \\
IRAS 10173+0828   & $-$     & $<0.29$ & $<0.34$ &   1.71  & $<0.33$ & $<0.32$ & 0.46    \\
IRAS 10339+1548   & 1.00    & 0.55    & $<0.34$ &   1.25  & 1.24    & $<1.13$ & 2.03    \\
IRAS 10378+1109   & 0.34    & $<0.25$ & $<0.35$ &   3.95  & $<0.61$ & $<0.63$ & 0.68    \\
IRAS 10485$-$1447 & $<1.14$ & $<0.20$ & $<0.23$ &   1.99  & $<0.44$ & $<0.48$ & 0.36    \\
IRAS 11028+3130   & $<1.06$ & $<0.70$ & $<0.28$ & $<0.63$ & $<0.44$ & $<0.38$ & $<0.74$ \\
IRAS 11180+1623   & $<0.80$ & $<0.70$ & $<0.36$ &   1.98  & $<0.50$ & $<0.39$ & 0.62    \\
IRAS 11524+1058   & $<0.78$ & $<0.56$ & $<0.35$ &   7.82  & $<0.43$ & $<0.37$ & $<0.77$ \\
IRAS 12018+1941   & $<0.27$ & $<0.28$ & $<0.59$ &   2.73  & $<0.52$ & $<0.38$ & 0.64    \\
IRAS 12032+1707   & $<1.33$ & 0.13    & $<0.48$ &   4.98  & 1.39:   & $<1.51$ & 1.57    \\
IRAS 12112+0305   & $-$     & 0.43    & $<0.88$ &  13.06  & $<0.76$ & $<0.55$ & 3.37    \\
IRAS 12540+5708   & $-$     & $<11.45$& $<4.40$ &  17.95  & $<8.05$ & $<6.86$ & 9.36    \\
IRAS 13218+0552   & $<0.66$ & $<0.54$ & $<0.46$ &   0.84  & $<0.71$ & $<1.23$ & $<1.24$ \\
IRAS 13428+5608   & $-$     & 7.69    & $<2.42$ &  41.21  & 10.62   & $<5.37$ & 29.05   \\
IRAS 13451+1232   & 0.57    & 1.57    & $<0.70$ &   4.71  & 0.71    & $<0.88$ & 4.73    \\
IRAS 14059+2000   & $<0.64$ & 0.35    & $<0.53$ &   2.83  & $<0.88$ & $<1.00$ & 2.65    \\
IRAS 14070+0525   & $<0.19$ & $<0.18$ & $<0.27$ &   1.33  & $<0.42$ & $<0.33$ & 2.01    \\
IRAS 14553+1245   & 0.47    & 0.71    & $<0.25$ &   3.24  & $<1.55$ & $<1.29$ & 3.10    \\
IRAS 15327+2340   & $-$     & $<0.79$ & $<3.60$ &  61.13  & $<6.29$ & $<5.29$ & 6.89    \\
IRAS 16090$-$0139 & 0.49    & 0.10    & $<0.49$ &   6.72  & $<0.65$ & $<0.76$ & 2.26    \\
IRAS 16255+2801   & 0.18    & 0.20    & $<0.27$ &   1.73  & $<0.82$ & $<0.68$ & 1.18    \\
IRAS 16300+1558   & 0.13    & $<0.15$ & $<0.28$ &   2.27  & $<0.33$ & $<0.25$ & 0.42    \\
IRAS 17207$-$0014 & $-$     & 0.40    & $<2.12$ &  38.84  & $<0.93$ & $<0.69$ & 8.42    \\
IRAS 18368+3549   & $<0.42$ & $<0.81$ & 0.11    &   6.91  & $<0.34$ & $<0.43$ & 1.07    \\
IRAS 18588+3517   & 0.57    & 0.37    & 0.24    &   5.12  & $<0.40$ & $<0.43$ & 1.87    \\
IRAS 20100$-$4156 & 0.33    & 0.23    & $<0.31$ &   6.73  & $<0.65$ & $<0.49$ & 1.71    \\
IRAS 20286+1846   & $<0.56$ & $<0.50$ & $<0.27$ &   1.67  & $<0.72$ & $<0.65$ & 0.37    \\
IRAS 21077+3358   & $<0.67$ & $<0.63$ & $<0.39$ &   3.06  & $<0.45$ & $<0.36$ & 1.09    \\
IRAS 21272+2514   & $<0.18$ & $<0.16$ & 0.16    &   2.29  & $<0.31$ & $<0.22$ & 0.35    \\
IRAS 22055+3024   & 0.12    & $<0.67$ & $<0.71$ &   4.55  & $<0.76$ & $<0.74$ & 1.05    \\
IRAS 22116+0437   & $<0.47$ & $<0.73$ & $<0.37$ &   2.25  & $<0.82$ & $<0.68$ & 1.33    \\
IRAS 22491$-$1808 & $-$     & 0.41    & $<0.54$ &   4.88  & $<0.55$ & $<0.43$ & 1.70    \\
IRAS 23028+0725   & $<0.55$ & $<0.61$ & $<0.39$ &   1.81  & $<0.90$ & $<0.72$ & 1.04    \\ 
IRAS 23233+0946   & 0.37    & 0.33    & $<0.43$ &   4.93  & $<0.43$ & $<0.38$ & 1.06    \\
IRAS 23365+3604   & $-$     & $<0.36$ & $<0.51$ &   8.51  & $<0.63$ & $<0.52$ & 1.12    \\
\hline    
IRAS 00163$-$1039 & $-$     & 3.40    & 0.43    & 80.95   & $<1.96$ & $<0.92$ & 14.53   \\
IRAS 01572+0009   & 0.73    & 3.06    & $<0.72$ & 6.15    & 5.51    & $<3.23$ & 10.34   \\
IRAS 05083+7936   & $-$     & $<1.19$ & $<1.57$ & 49.40   & 0.60    & 0.48    & 7.63    \\
IRAS 06538+4628   & $-$     & 0.80    & 1.09    & 47.39   & 0.69    & $<1.16$ & 6.07    \\
IRAS 08559+1053   & $<0.37$ & 0.56    & $<0.43$ & 8.38    & 0.51    & 0.35    & 1.87    \\
IRAS 09437+0317   & $-$     & $<1.04$ & $<0.61$ & 8.74    & $<0.64$ & $<0.51$ & 1.22    \\
IRAS 10565+2448   & $-$     & $<0.76$ & $<1.34$ & 57.60   & $<1.29$ & 0.67    & 7.65    \\
IRAS 11119+3257   & $<0.83$ & 0.31    & $<0.78$ & 2.19    & $<0.71$ & $<0.58$ & 1.89    \\
IRAS 13349+2438   & 2.76    & 1.66    & $<0.45$ & 1.43    & 0.81    & $<0.99$ & 3.50    \\
IRAS 15001+1433   & 0.39    & 0.28    & $<0.30$ & 6.61    & 1.08    & $<0.98$ & 2.62    \\
IRAS 15206+3342   & 2.01    & 3.82    & 0.23    & 10.96   & $<0.34$ & $<0.41$ & 19.87   \\
IRAS 20460+1925   & $<0.39$ & $<0.66$ & $<0.42$ & $<0.44$ & $<0.60$ & $<0.51$ & $<0.69$ \\
IRAS 23007+0836   & $-$     & 8.67    & $<1.84$ & 179.04  & 8.36    & $<4.23$ & 33.69   \\
IRAS 23394$-$0353 & $-$     & 0.72    & 0.57    & 46.75   & $<1.45$ & $<1.08$ & 7.71    \\
IRAS 23498+2423   & 0.23    & 1.27    & $<0.23$ & 3.10    & 0.92    & $<0.91$ & 7.79    \\
\enddata
\tablecomments{Line fluxes are given in 10$^{-21}$ W cm$^{-2}$. $-$ indicates that the redshifted line wavelength lay outside the range of the IRS.}
\end{deluxetable*}

\begin{deluxetable*}{lrrrrrrr}
\tabletypesize{\scriptsize}
\tablecaption{Atomic line fluxes for high-resolution spectra: $17-35$~microns\label{tbl-hrlines2}}
\tablewidth{0pt}
\tablehead{
\colhead{Object} & 
\colhead{[Fe \tiny{II}\scriptsize]} &
\colhead{[S \tiny{III}\scriptsize]} &
\colhead{[Ne \tiny{V}\scriptsize]} &
\colhead{[O \tiny{IV}\scriptsize]} &
\colhead{[Fe \tiny{II}\scriptsize]} &
\colhead{[S \tiny{III}\scriptsize]} &
\colhead{[Si \tiny{II}\scriptsize]} 
\\
\colhead{$\lambda_{rest}$ [\um]} & 
\colhead{17.936} & 
\colhead{18.713} &
\colhead{24.318} &
\colhead{25.890} &
\colhead{25.988} &
\colhead{33.481} &
\colhead{34.815}
}
\hline    
\startdata
IRAS 01355$-$1814 									& $<0.51$ & $<1.52$ & $<0.78$   & $<1.28$   & $<0.82$  & $-$       & $-$       \\
IRAS 01418+1651   									& $<0.67$ & 0.94    & $<3.84$   & $<3.52$   & $<4.05$  & $<24.62$  & 19.43     \\
IRAS 01562+2528   									& $<0.77$ & 0.67    & $<0.73$   & $<2.06$   & $<1.29$  & $-$       & $-$       \\
IRAS 02524+2046   									& $<1.11$ & 0.47    & $<1.04$   & $<0.96$   & $<0.59$  & $-$       & $-$       \\
IRAS 03521+0028   									& 1.70    & 1.18    & $<1.02$   & $<0.93$   & $<0.59$  & $-$       & $-$       \\
IRAS 04121+0223   									& $<0.92$ & 0.93    & $<1.06$   & 0.97      & $<1.34$  & $-$       & $-$       \\
IRAS 04454$-$4838 									& $<0.27$ & $<1.00$ & $<7.75$   & $<9.59$   & $<6.29$  & 0.90      & $<87.49$  \\
IRAS 06487+2208   									& $<0.87$ & 4.89    & $<1.36$   & 0.48      & 1.31     & $-$       & $-$       \\
IRAS 07163+0817   									& $<1.04$ & 0.81    & $<1.73$   & $<0.78$   & $<0.71$  & $-$       & $-$       \\
IRAS 07572+0533   									& $<0.61$ & $<1.15$ & $<1.37$   & $<1.53$   & $<1.03$  & $-$       & $-$       \\
IRAS 08201+2801   									& $<0.84$ & 0.41    & $<0.50$   & $<0.66$   & $<0.50$  & $-$       & $-$       \\
IRAS 08449+2332   									& $<0.60$ & 1.62    & $<0.73$   & $<0.69$   & $<0.40$  & $-$       & $-$       \\
IRAS 08474+1813   									& $<1.09$ & 0.86    & $<1.27$   & $<3.04$   & $<1.86$  & $-$       & $-$       \\
IRAS 09039+0503   									& $<0.85$ & 0.88    & $<0.55$   & $<1.11$   & 0.48     & $-$       & $-$       \\
IRAS 09539+0857   									& $<1.82$ & $<1.02$ & $<1.41$   & $<2.05$   & $<1.36$  & $-$       & $-$       \\
IRAS 10035+2740   									& $<0.97$ & 0.78    & $<0.95$   & $<2.11$   & $<1.37$  & $-$       & $-$       \\
IRAS 10039$-$3338 									& $<0.81$ & 6.57    & $<14.49$  & $<11.09$  & $<7.54$  & $<10.49$  & $101.40$  \\
IRAS 10173+0828   									& $<0.42$ & $<1.36$ & $<1.33$   & $<2.42$   & $<1.61$  & 2.55      & 81.70     \\
IRAS 10339+1548   									& 0.49    & 1.17    & 0.52      & 2.85      & 0.89     & $-$       & $-$       \\
IRAS 10378+1109   									& $<0.95$ & 1.82    & $<0.87$   & $<1.99$   & $<1.30$  & $-$       & $-$       \\
IRAS 10485$-$1447 									& $<0.87$ & $<1.08$ & $<0.91$   & $<1.08$   & $<0.66$  & $-$       & $-$       \\
IRAS 11028+3130   									& $<0.47$ & $<0.71$ & $<3.07$   & $<4.45$   & $<3.14$  & $-$       & $-$       \\
IRAS 11180+1623   									& $<0.71$ & 0.89    & $<0.85$   & $<1.00$   & $<0.69$  & $-$       & $-$       \\
IRAS 11524+1058   									& $<0.52$ & $<1.61$ & $<1.49$   & $<0.93$   & $<0.70$  & $-$       & $-$       \\
IRAS 12018+1941   									& $<1.01$ & $<1.32$ & $<1.22$   & $<1.71$   & $<1.30$  & $-$       & $-$       \\
IRAS 12032+1707   									& $<1.22$ & 1.18    & 0.66      & $<1.50$   & $<1.00$  & $-$       & $-$       \\
IRAS 12112+0305   									& 0.68    & 4.32    & $<1.50$   & $<7.56$   & $<4.65$  & 8.77      & $-$       \\
IRAS 12540+5708   									& $<4.40$ & $<5.05$ & $<22.99$ & $<18.84$  & $<14.43$ & $<49.65$   & $<163.66$ \\
IRAS 13218+0552   									& $<0.78$ & $<2.27$ & $<1.85$   & $<2.93$   & $<1.76$  & $-$       & $-$       \\
IRAS 13428+5608   									& $<1.32$ & 16.67   & 9.28      & 75.13     & $<29.79$ & 28.14     & $<134.61$ \\
IRAS 13451+1232   									& $<1.19$ & 1.15    & $<2.50$   & $<5.17$   & $<1.85$  & $-$       & $-$       \\
IRAS 14059+2000   									& $<0.97$ & 1.68    & $<1.04$   & $<1.70$   & $<1.11$  & $-$       & $-$       \\
IRAS 14070+0525   									& $<0.53$ & $<0.77$ & $<1.25$   & $<1.98$   & $<1.37$  & $-$       & $-$       \\
IRAS 14553+1245   									& $<0.89$ & 1.58    & $<0.89$   & $<2.38$   & $<1.49$  & $-$       & $-$       \\
IRAS 15327+2340   									& $<1.04$ & 5.19    & $<24.96$  & $<24.33$  & $<15.22$ & $<151.68$ & $<202.18$ \\
IRAS 16090$-$0139 									& 1.03    & 3.07    & $<3.95$   & 1.12      & $<2.23$  & $-$       & $-$       \\
IRAS 16255+2801   									& $<5.36$ & 1.54    & $<12.06$  & $<7.16$   & $<4.83$  & $-$       & $-$       \\
IRAS 16300+1558   									& $<1.20$ & 0.50    & $<1.23$   & $<0.88$   & $<0.70$  & $-$       & $-$       \\
IRAS 17207$-$0014 									& $<0.51$ & 6.39    & $<6.40$   & $<11.27$  & $<7.70$  & 12.24     & 46.91     \\
IRAS 18368+3549   									& $<0.49$ & 1.18    & $<1.07$   & $<0.91$   & $<0.59$  & $-$       & $-$       \\
IRAS 18588+3517   									& $<3.75$ & 7.96    & $<4.91$   & $<4.13$   & $<4.05$  & $-$       & $-$       \\
IRAS 20100$-$4156 									& $<0.69$ & 2.86    & $<3.58$   & 1.20      & 1.34     & $-$       & $-$       \\
IRAS 20286+1846   									& $<0.30$ & 0.44    & $<0.58$   & $<0.73$   & $<0.70$  & $-$       & $-$       \\
IRAS 21077+3358   									& $<0.50$ & 0.77    & $<0.80$   & $<0.60$   & $<0.47$  & $-$       & $-$       \\
IRAS 21272+2514   									& $<0.21$ & 0.45    & $<0.50$   & $<0.68$   & $<0.46$  & $-$       & $-$       \\
IRAS 22055+3024   									& $<0.51$ & 1.04    & $<2.12$   & $<2.30$   & $<1.34$  & $-$       & $-$       \\
IRAS 22116+0437   									& $<1.07$ & 1.04    & $<2.16$   & $<2.54$   & $<2.35$  & $-$       & $-$       \\
IRAS 22491$-$1808 									& $<0.72$ & 1.86    & $<3.40$   & $<7.11$   & $<3.35$  & 13.72     & $-$       \\
IRAS 23028+0725   									& $<1.05$ & $<1.15$ & $<1.72$   & $<2.70$   & $<1.72$  & $-$       & $-$       \\ 
IRAS 23233+0946   									& $<0.89$ & 3.13    & $<1.06$   & 1.20      & 0.66     & $-$       & $-$       \\
IRAS 23365+3604   									& $<0.44$ & 4.29    & $<4.74$   & $<11.96$  & $<7.97$  & 6.81      & $-$       \\
\hline    
IRAS 00163$-$1039 									& $<1.70$ & 30.85   & $<7.59$   & 1.42      & 3.35     & 34.70     & 73.84     \\
IRAS 01572+0009   									& $<0.89$ & 1.66    & 4.76      & 9.82      & $<6.12$  & $-$       & $-$       \\
IRAS 05083+7936   									& $<1.19$ & 18.67   & $<1.89$   & 1.28      & 1.33     & 28.00     & 71.27     \\
IRAS 06538+4628   									& 1.17    & 19.11   & $<2.33$   & $<6.97$   & 2.70     & 37.60     & 57.37     \\
IRAS 08559+1053   									& $<0.93$ & 1.35    & $<1.34$   & 2.54      & 0.45     & $-$       & $-$       \\
IRAS 09437+0317   									& $<0.70$ & 3.33    & $<0.99$   & 0.47      & 0.68     &  9.45     & 23.27     \\
IRAS 10565+2448   									& $<1.02$ & 12.42   & $<3.07$   & $<5.13$   & $<2.35$  & 20.89     & 51.37     \\
IRAS 11119+3257   									& $<2.30$ & $<1.75$ & $<2.40$   & $<3.08$   & $<1.95$  & $-$       & $-$       \\
IRAS 13349+2438   									& $<4.54$ & 1.68    & 3.46      & 7.28      & $<5.11$  & $-$       & $-$       \\
IRAS 15001+1433   									& $<0.78$ & 2.36    & 0.67      & 1.21      & 0.56     & $-$       & $-$       \\
IRAS 15206+3342   									& $<1.03$ & 8.58    & 1.36      & 0.74      & 1.21     & $-$       & $-$       \\
IRAS 20460+1925   									& $<0.72$ & $<0.79$ & $<3.68$   & 1.95      & $<1.01$  & $-$       & $-$       \\
IRAS 23007+0836   									& $<3.28$ & 70.16   & 15.51     & 30.98     & 9.47     & 97.56     & 188.36    \\
IRAS 23394$-$0353 									& $<1.10$ & 17.16   & $<1.92$   & 1.27      & 2.43     & 44.29     & 53.48     \\
IRAS 23498+2423   									& $<0.55$ & 1.13    & 1.26      & 4.61      & $<3.77$  & $-$       & $-$       \\
\enddata
\tablecomments{Line fluxes are given in 10$^{-21}$ W cm$^{-2}$. $-$ indicates that the redshifted line wavelength lay outside the range of the IRS.}
\end{deluxetable*}

The IRS is designed to make accurate measurements of narrow atomic transitions in the SH and LH modules; however, several lines are also detected in the LR modules, most often the powerful neon, sulfur, and \htwo~transitions. The only emission lines visible in the low-resolution modules without a corresponding detection in high-resolution are the \htwo~S(7) line at 5.5~\um~and the H$_2$ S(5)/\arII~complex near 6.7~\um; this is due to the SH low-wavelength cutoff at $\lambda_{rest}=8.2-9.0$~\um, depending on the redshift of the galaxy. All other atomic emission features observed in the low-resolution modules have corresponding detections in high-resolution; furthermore, blending of narrow lines makes accurate measurements of flux difficult in the LR modules. For isolated lines with well-defined surrounding continuum, our fluxes are consistent for measurements in both low- and high-resolution. 

We note the presence of two features which have no obvious identifications occurring in the LH spectra for multiple objects: one is an emission feature seen near 29~\um~(a prominent example occurs for IRAS~17539+2935) and the second is an absorption feature near 30~\um. The two are often paired and are seen in $\sim50\%$ of the galaxies observed. The \emph{rest} wavelengths of the transitions, however, vary significantly from object to object (with a standard deviation of $\sigma_\lambda \simeq 0.7$~\um), while the \emph{observed} wavelengths are nearly fixed ($\sigma_\lambda \lesssim 0.05$~\um). This implies that the features are either artifacts of the extraction process or that both the emission and absorption come from unidentified foreground features with little to no Doppler shift. Given that we see no evidence for these unidentified lines in any of the LR spectra (which should be detectable, given the high S/N for many of the features), we consider both to be spurious. 

\subsection{Molecular hydrogen}\label{ssec-h2}

We detected multiple emission lines from the pure rotational series of molecular hydrogen in both OHMs and non-masing galaxies. At redshifts of $z\lesssim0.1$, transitions from \htwo~S(0) at 28.22~\um~to \htwo~S(3) at 9.67~\um~are visible in the HR modules; in addition, the LR module is capable of detecting lines as far out as the S(7) transition at 5.51~\um. In each case, the line number [{\it eg}, 0 for \htwo~S(0)] indicates the rotational quantum number of the lower state ($J=2\rightarrow0$) for the quadrupole S-branch transition. $\Delta J=2$ results in two separate branches: ortho (parallel nuclear spin, odd $J$) and para (anti-parallel nuclear spin, even $J$).

We detected at least one \htwo~line in 49/51 OHMs and 13/15 non-masing galaxies, with S(1) seen in all objects for which at least one molecular hydrogen transition is reported. The higher-order S(2) and S(3) lines are seen in roughly 2/3 of the sample, while the para ground state S(0) transition is detected in only $\sim15\%$ of the sample. Our line detection rate is consistent with results from the SINGS galaxies examined in \citet{rou07} and the ULIRG sample of \citet{hig06}. Lower detection rates of S(0) and S(2) are likely due to a combination of the intrinsic ortho-para ratio as well as rising continuum levels near 28~\um~that can obscure weak line emission by S(0). \citet{hig06} find that the S(2)/S(3) ratios are consistent with no significant differential extinction for the two lines, which is supported by numerous detections in our sample of S(3) line emission superimposed on optically deep silicate absorption near 9.7~\um. We thus applied no extinction or reddening corrections to the line fluxes (Table~\ref{tbl-h2}).  

\begin{deluxetable*}{lrrrrrrrcrrc}
\tabletypesize{\scriptsize}
\tablecaption{Molecular H$_2$ gas properties\label{tbl-h2}}
\tablewidth{0pt}
\tablehead{
\colhead{Object} & 
\colhead{H$_2$~S(7)} &
\colhead{H$_2$~S(5)} &
\colhead{H$_2$~S(4)} &
\colhead{H$_2$~S(3)} &
\colhead{H$_2$~S(2)} &
\colhead{H$_2$~S(1)} &
\colhead{H$_2$~S(0)} &
\colhead{$T_{warm}$} &
\colhead{$T_{hot}$} &
\colhead{$M_{warm}$} &
\colhead{$M_{hot}$}
\\
\colhead{$\lambda_{rest}$ [\um]} & 
\colhead{5.51~\um} & 
\colhead{6.91~\um} & 
\colhead{8.03~\um} & 
\colhead{9.67~\um} & 
\colhead{12.28~\um} & 
\colhead{17.04~\um} & 
\colhead{28.22~\um} & 
\colhead{[K]} & 
\colhead{[K]} & 
\colhead{[$10^7$ M$_\sun$]} & 
\colhead{[$10^7$ M$_\sun$]} 
}
\startdata
IRAS 01355$-$1814 & $<1.52$  & $<3.25$   & $-$     & 0.22    & 0.63    & 1.13    & $<1.63$ &  262 &      &  4.74 &       \\
IRAS 01418+1651   & $<5.29$  & 6.96:     & $-$     & 1.16    & 0.94    & 2.15    & $<5.40$ &  320 &      &  0.14 &       \\
IRAS 01562+2528   & $<3.25$  & $<7.06$   & $-$     & $<0.52$ & $<0.39$ & 0.94    & $<0.92$ &      &      &       &       \\
IRAS 02524+2046   & $<1.52$  & 0.88:     & $-$     & $<2.51$ & $<0.46$ & 0.88    & $<1.03$ &      &      &       &       \\
IRAS 03521+0028   & $<2.89$  & 0.58:     & $-$     & 0.63    & 0.47    & 1.69    & $<0.55$ &  292 &      &  4.24 &       \\
IRAS 04121+0223   & $<3.54$  & 1.03:     & $-$     & $<1.62$ & $<0.43$ & $<0.73$ & $<1.06$ &      &      &       &       \\
IRAS 04454$-$4838 & $<3.31$  & $<13.24$  & $-$     & 1.04    & 1.27    & 3.05    & 0.76    &  222 &      &  0.82 &       \\
IRAS 06487+2208   & $<5.99$  & 0.97:     & $-$     & 2.10    & 1.02    & 1.80    & $<1.52$ &  381 &      &  4.13 &       \\
IRAS 07163+0817   & $<1.84$  & $<4.58$   & $-$     & $<0.77$ & 0.04    & 0.47    & 1.68    &  110 &      &  0.60 &       \\
IRAS 07572+0533   & $<1.03$  & $<2.62$   & $-$     & $<0.90$ & $<0.50$ & $<0.99$ & $<3.41$ &      &      &       &       \\
IRAS 08201+2801   & $<3.41$  & $<10.84$  & $-$     & 0.28    & $<0.90$ & 0.56    & $<0.91$ &  313 &      &  1.79 &       \\
IRAS 08449+2332   & $<3.67$  & 1.22:     & $-$     & 0.63    & 0.33    & 1.36    & $<0.70$ &  301 &      &  3.48 &       \\
IRAS 08474+1813   & $<1.07$  & 0.61:     & $-$     & 0.75    & $<0.46$ & 0.27    & $<1.33$ &  493 &      &  0.68 &       \\
IRAS 09039+0503   & 1.13     & 2.69:     & $-$     & 2.21    & 1.46    & 2.91    & 1.27    &  198 & 978  &  4.88 & 0.42  \\
IRAS 09539+0857   & $<2.40$  & $<6.29$   & $-$     & 0.53    & 0.26    & 1.07    & $<0.95$ &  308 &      &  1.93 &       \\
IRAS 10035+2740   & $<0.98$  & 0.81:     & $-$     & 0.70    & $<0.63$ & 1.06    & $<2.31$ &  334 &      &  3.29 &       \\
IRAS 10039$-$3338 & $<28.84$ & $<46.95$  & $-$     & 3.27    & 1.71    & 3.95    & $<7.46$ &  348 &      &  0.46 &       \\
IRAS 10173+0828   & $<5.73$  & 3.00:     & $-$     & 0.58    & 0.49    & 1.38    & $<1.67$ &  302 &      &  0.33 &       \\
IRAS 10339+1548   & $<1.09$  & $<2.88$   & $-$     & $<0.58$ & $<0.42$ & 0.42    & $<1.94$ &      &      &       &       \\
IRAS 10378+1109   & 0.53     & 1.42:     & $-$     & 1.82    & 0.55    & 2.25    & $<0.61$ &  315 & 876  &  4.59 & 0.46  \\
IRAS 10485$-$1447 & 0.52     & $<5.57$   & $-$     & 0.18    & 0.14    & 0.28    & $<0.76$ &  307 & 1540 &  0.55 & 0.03  \\
IRAS 11028+3130   & $<0.44$  & $<1.83$   & $-$     & $<0.61$ & $<0.36$ & 0.39    & $<1.75$ &      &      &       &       \\			
IRAS 11180+1623   & $<2.25$  & 0.51:     & $-$     & 0.39    & 0.61    & 1.06    & $<1.88$ &  302 &      &  3.32 &       \\
IRAS 11524+1058   & $<1.99$  & $<4.35$   & $-$     & $<0.43$ & $<0.42$ & 0.66    & $<2.14$ &      &      &       &       \\
IRAS 12018+1941   & $<3.27$  & $<8.84$   & $-$     & 0.47    & 0.36    & 1.28    & 2.21    &  165 &      &  4.10 &       \\
IRAS 12032+1707   & $<2.78$  & 4.93:     & $<0.27$ & 0.81    & 0.58    & 1.61    & $<3.34$ &  312 &      &  9.24 &       \\
IRAS 12112+0305   & 7.03     & 5.89:     & $-$     & 1.87    & 1.66    & 3.71    & 1.30    &  195 & 1680 &  2.01 & 0.08  \\
IRAS 12540+5708   & $<32.27$ & $<49.24$  & $-$     & 2.42    & 3.21    & 6.23    & $<17.62$&  301 &      &  1.07 &       \\
IRAS 13218+0552   & $<4.32$  & $<7.55$   & $-$     & 0.32    & 0.42    & 0.97    & $<2.91$ &  292 &      &  4.83 &       \\
IRAS 13428+5608   & $<13.14$ & 8.96:     & $-$     & 7.77    & 4.83    & 8.63    & $<9.82$ &  360 &      &  1.19 &      \\
IRAS 13451+1232   & $<8.46$  & 3.26:     & $-$     & 1.65    & 1.10    & 2.61    & $<0.79$ &  329 &      &  4.18 &       \\
IRAS 14059+2000   & 0.91     & 1.55:     & $-$     & 2.12    & 0.73    & 2.36    & $<0.88$ &  319 & 944  &  3.90 & 0.40  \\
IRAS 14070+0525   & $<2.54$  & $<5.85$   & $<0.41$ & 0.28    & 0.22    & 1.30    & $<0.85$ &  260 &      & 11.42 &       \\
IRAS 14553+1245   & 0.32     & $<9.92$   & $-$     & 0.92    & 0.19    & 1.05    & $<0.82$ &  317 & 906  &  1.76 & 0.19  \\
IRAS 15327+2340   & 22.70    & 41.6:     & $-$     & $<0.56$ & 7.66    & 13.68   & 14.53   &  159 &      &  0.42 &       \\
IRAS 16090$-$0139 & 1.20     & $<15.12$  & $-$     & 1.10    & $<0.28$ & 2.12    & $<2.93$ &  299 & 1160 &  4.09 & 0.20  \\
IRAS 16255+2801   & $<1.82$  & 0.21:     & $-$     & $<0.59$ & 0.28    & 0.55    & $<4.18$ &  348 &      &  1.07 &       \\
IRAS 16300+1558   & $<1.77$  & $<5.63$   & 0.31    & 0.56    & 0.42    & 1.58    & $<1.31$ &  287 &      & 11.33 &       \\
IRAS 17207$-$0014 & 6.12     & 10.2:     & $-$     & 4.56    & 4.44    & 7.51    & $<3.60$ &  311 & 1230 &  1.30 & 0.07 \\
IRAS 18368+3549   & $<6.98$  & 4.22:     & $-$     & 1.11    & 0.70    & 1.38    & $<0.81$ &  349 &      &  1.95 &       \\
IRAS 18588+3517   & $<7.44$  & $<15.54$  & $-$     & 0.92    & 0.76    & 1.61    & $<5.12$ &  324 &      &  1.89 &       \\
IRAS 20100$-$4156 & $<9.04$  & $<15.77$  & $-$     & 0.78    & 0.35    & 0.92    & 0.94    &  195 &      &  1.61 &       \\
IRAS 20286+1846   & $<1.74$  & 0.35:     & $-$     & $<0.78$ & 0.30    & 0.92    & $<0.51$ &  282 &      &  1.80 &       \\
IRAS 21077+3358   & $<2.13$  & 1.59:     & $-$     & 0.67    & 0.34    & 1.16    & $<1.29$ &  319 &      &  4.05 &       \\
IRAS 21272+2514   & 0.39     & 0.51:     & $-$     & 0.36    & 0.28    & 0.91    & $<0.46$ &  287 & 1160 &  2.22 & 0.09  \\
IRAS 22055+3024   & 0.65     & 2.52:     & $-$     & 1.30    & 1.09    & 1.61    & $<1.39$ &  320 & 976  &  2.72 & 0.24  \\
IRAS 22116+0437   & $<2.45$  & $<6.21$   & $-$     & 0.74    & 0.64    & 1.09    & $<1.82$ &  340 &      &  4.70 &       \\
IRAS 22491$-$1808 & $<7.01$  & $<14.81$  & $-$     & 0.76    & 0.92    & 2.10    & $<8.22$ &  298 &      &  1.23 &       \\
IRAS 23028+0725   & $-$      & $-$       & $-$     & 0.48    & 0.45    & 0.77    & $<1.67$ &  333 &      &  1.86 &       \\
IRAS 23233+0946   & $<3.20$  & 1.60:     & $-$     & 1.03    & 0.80    & 1.47    & $<0.73$ &  340 &      &  2.53 &       \\
IRAS 23365+3604   & $<7.02$  & 6.98:     & $-$     & 1.26    & 0.73    & 2.14    & $<5.75$ &  321 &      &  0.84 &      \\
\hline                       	                                                                                        
IRAS 00163$-$1039 & $<9.00$  & 24.46:    & $-$     & 2.79    & 2.87    & 5.13    & $<2.63$ &  326 &      &  0.32 &       \\
IRAS 01572+0009   & 2.15     & 1.21:     & $-$     & 0.60    & $<0.71$ & 2.15    & $<1.36$ &  268 & 1650 &  6.26 & 0.15  \\
IRAS 05083+7936   & $<8.11$  & 27.31:    & $-$     & 2.34    & 2.73    & 3.59    & $<2.75$ &  341 &      &  0.99 &       \\
IRAS 06538+4628   & $<3.69$  & 14.56:    & $-$     & $<0.63$ & 3.28    & 8.86    & 2.63    &  189 &      &  0.38 &       \\
IRAS 08559+1053   & $<5.91$  & 1.76:     & $-$     & 0.72    & 0.64    & 1.83    & $<0.74$ &  298 &      &  4.44 &       \\
IRAS 09437+0317   & $<9.39$  & 22.60:    & $-$     & $<0.36$ & 0.68    & 2.68    & 2.20    &  153 &      &  0.11 &       \\
IRAS 10565+2448   & $<12.33$ & 32.87:    & $-$     & 3.34    & 1.95    & 5.73    & $<3.77$ &  320 &      &  1.06 &       \\
IRAS 11119+3257   & $<6.58$  & $<7.87$   & $-$     & 0.42    & $<1.02$ & 2.47    & $<2.23$ &  256 &      & 10.21 &       \\
IRAS 13349+2438   & $<23.79$ & $<20.52$  & $-$     & $<1.67$ & $<0.83$ & $<1.05$ & $<1.62$ &      &      &       &       \\
IRAS 15001+1433   & $<4.21$  & 1.72:     & $-$     & 0.44    & 0.24    & 1.25    & $<0.84$ &  283 &      &  3.72 &       \\
IRAS 15206+3342   & $<8.31$  & 1.61:     & $-$     & 0.65    & 0.46    & 0.94    & 1.02    &  196 &      &  1.54 &       \\
IRAS 20460+1925   & $-$      & $-$       & $-$     & $<0.48$ & $<0.43$ & $<0.88$ & $<4.24$ &      &      &       &       \\
IRAS 23007+0836   & $<23.20$ & 77.35:    & $-$     & $<3.53$ & 6.30    & 12.90   & $<6.28$ &  342 &      &  0.27 &      \\
IRAS 23394$-$0353 & $<14.60$ & 32.52:    & $-$     & 3.15    & 2.44    & 5.24    & 1.79    &  226 &      &  0.23 &       \\
IRAS 23498+2423   & $<1.25$  & 0.49:     & $-$     & 0.34    & $<0.29$ & 0.89    & $<1.50$ &  298 &      &  4.66 &       \\
\enddata
\tablecomments{Line fluxes are given in $10^{-21}$ W cm$^{-2}$. $-$ indicates that the redshifted line lay outside of the IRS spectral range. Fluxes for \htwo~S(7) and S(5) are measured in the SL module; all other lines are measured in the SH and LH modules. S(5) lines are tentative upper limits (indicated by a :) due to possible blending with [Ar~\tiny{II}\scriptsize]~at 6.99~\um; see \S\ref{ssec-h2}. }
\end{deluxetable*}

We did not detect the S(6) line in any galaxy, while S(4) showed a single detection in IRAS~16300+1558. Since these lines are excited by hotter gas ($T\sim1000$~K), they are typically weaker than the lower states which probe the larger reservoir of cool gas. In addition, both lines are only visible in the low-resolution modules at $z\sim0.1$. This means that deblending is a significant issue, since both lines lie near broad PAH emission complexes. 11 OHMs and one non-masing galaxy show the unresolved S(7) ortho line at 5.51~\um~in the SL module. 

Measurement of the S(5) line presents a particular problem due to its location in a crowded section of the spectra. Its rest wavelength of 6.91~\um~lies near the \arII~feature at 6.99~\um; in addition, both features are bracketed by possible hydrocarbon absorption at 6.85 and 7.25~\um. This not only creates difficulties in establishing a reliable continuum, but also in deblending the \arII~and the S(5) emission (see \S\ref{sssec-hac}). Emission in the \arII/\htwo~S(5) complex is seen in more than half of our sample, however, and so we present measurements for the \emph{entire} feature, including blended emission from both lines. We caution that these fluxes should be viewed as upper limits for either \arII~or \htwo~S(5) emission, since the SL module does not have sufficient resolution to separate the two features. 

For galaxies in which multiple \htwo~lines are observed, we fit excitation temperatures ($T_{ex}$) to the molecular gas following the methods of \citet{rig02} and \citet{hig06}. We assume that the emission is optically thin (so that the lines are unsaturated), populations are in local thermodynamic equilibrium (LTE), and that the sources are unresolved in the {\it Spitzer} beam. The luminosity of a molecular emission line for the transition from $(J+2)\rightarrow J$ is then $L_J = A_J \times \Delta E_J \times N_{J+2}$, where $A_J$ is the Einstein-A coefficient, $\Delta E_J$ is the energy of the transition, and $N_{J+2}$ is the number of molecules in the $J+2$ state. The partition function for a given symmetry branch is:

\begin{equation}
\label{eqn-linelum3}
Z_{J_{o/p}} = \sum_{J_{o/p}} g_{J}~exp[-E_{J} / k T_{ex}]
\end{equation}

\noindent where $T_{ex}$ is the excitation temperature and we sum only over a single symmetry branch (ortho or para). The statistical weights are $g_J=(2J+1)\times J_s$, where $J_s=1$ for the para branch (even $J$) and $J_s=3$ for ortho (odd $J$). 

\begin{figure*}
\includegraphics[scale=1.0]{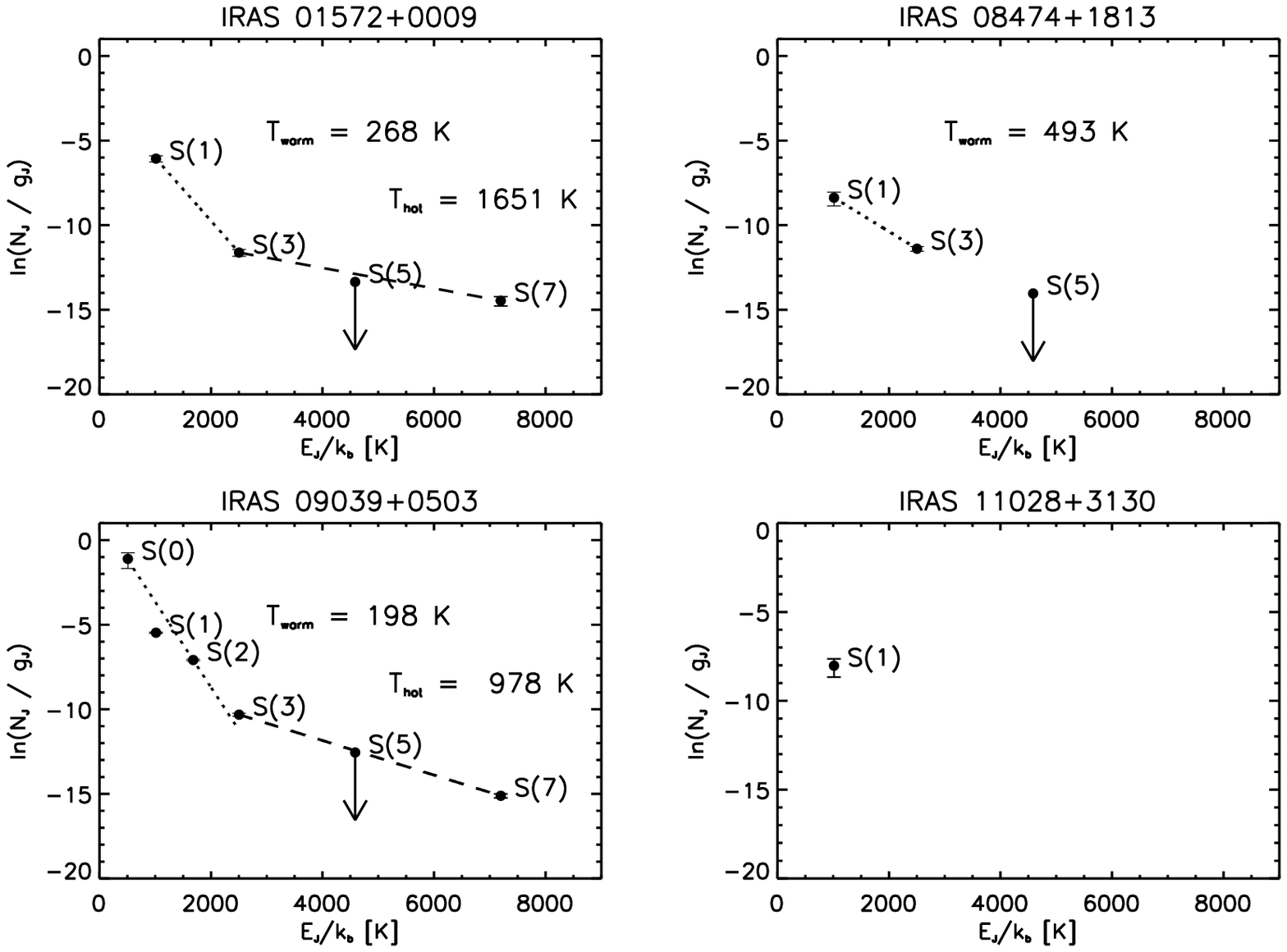}
\caption{Example \htwo~excitation diagrams for both non-masing (IRAS~01572+0009) and OHM galaxies (all others). Two galaxies ({\it left}) are fit with both warm and hot excitation temperatures; IRAS~08474+1813 fits only a warm component since the higher~$J$ lines are not detected. IRAS~11028+3130 shows an example of a galaxy with only a single \htwo~detection (for which no $T_{ex}$ can be determined). Dotted lines are fit to the warm gas for all detections from S(0) to S(3); dashed lines are fit to the hotter gas using detections of S(3), S(4) and S(7). The S(5) line is always an upper limit due to possible blending from \arII~and is not used in the temperature fits.\label{fig-h2temp_2x2}}
\end{figure*}

Assuming that the lines are in LTE, the ratio of level populations follows a Boltzmann distribution such that $N_J\propto~g_J~exp[-E_J/k~T_{ex}]$. The inverse slope of the best-fit line of an excitation diagram yields $T_{ex}$ - Figure~\ref{fig-h2temp_2x2} shows examples of temperature fits for our data. The total warm \htwo~mass can be then calculated from $T_{ex}$ and the flux ($F_J$) from any transition as:

\begin{equation}
\label{eqn-linelum4}
M_{tot} = m_{H_2} \times \phi_{o/p} \times \frac{(4 \pi D_L^2) F_J Z_{J_{o/p}}}{A_J \Delta E_J g_J~exp[-E_J/kT_{ex}]}
\end{equation}

\noindent where $\phi_{o/p}$ is a numerical factor accounting for the ortho-to-para ratio (assumed to be 3:1), $m_{H_2}$ is the mass of the hydrogen molecule, and $D_L$ is the luminosity distance. 

For cases where the S(7) line was detected, a single excitation temperature gives a poor fit to the full set of transitions. In these cases, we first fit $T_{ex}$ between S(3) and S(7), measuring hotter gas. We then subtracted this component from the S(0) to S(3) fluxes, and fit a second $T_{ex}$ to the warm gas component. This decreased the mean warm $T_{ex}$ by $\sim20$~K, with a negligible effect on the gas mass. We calculated the warm \htwo~mass using the flux in the S(1) transition and the hot gas mass using the S(3) flux (Table~\ref{tbl-h2}). 

	Both \cite{hig06} and \cite{rou07} suggest that the \htwo~emission arises from far-ultraviolet photons from massive stars powering photodissociation regions (PDRs). Detections of the \htwo~S(3) transition in nearly all objects implies that the silicate absorption at 9.7~\um~must be partially background to the warm molecular gas seen in emission. Since the dust is very optically thick in almost all ULIRGs, this means that at least some molecular gas (and possibly other atomic transitions) actually come from superficial layers at the edge of the merging system. Given that the OHM is typically formed within the central kiloparsec of the host galaxy, a link between the observed warm \htwo~gas and the OHM is uncertain. 
	
	11 objects in the OHM sample and 2 non-masing galaxies have CO detections published in the literature \citep{sol97,gao04a}. The beam width used for CO observations is several times that of the HR slits; since most ULIRGs are unresolved in the {\it Spitzer} beam, we consider the gas mass estimates to be comparable. The cold gas masses derived using a ULIRG-calibrated $M_{H_2}/L_{CO}$ ratio of $\sim1.4 M_\sun$/(K km s$^{-1}$ pc$^2$) give a warm gas mass fraction for the OHMs ranging from $0.04-0.8\%$, with the gas fraction of the non-masing galaxies lying in a similar range ($0.06-0.1\%$). This is comparable to warm gas fractions in ULIRGs from \citet{hig06}, implying that the mid-IR \htwo~lines probe only a small amount of the total gas mass in these galaxies. The bulk of the remaining portion is likely cold gas without sufficient energy to excite rotational transitions in the mid-IR. 

\subsection{PAH emission}\label{ssec-pah}

\begin{deluxetable*}{lrrrrrrrrrr}
\tabletypesize{\scriptsize}
\tablecaption{PAH emission features in low-resolution spectra\label{tbl-pah}}
\tablewidth{0pt}
\tablehead{
\colhead{} & 
\multicolumn{2}{c}{\underline{PAHFIT luminosity}} &
\multicolumn{3}{c}{\underline{PAHFIT EW}} &
\multicolumn{2}{c}{\underline{Spline-fit luminosity}} &
\multicolumn{3}{c}{\underline{Spline-fit EW}}
\\
\colhead{Object} & 
\colhead{6.2} &
\colhead{11.3} & 
\colhead{6.2} &
\colhead{6.2 ice} &
\colhead{11.3} &
\colhead{6.2} &
\colhead{11.3} & 
\colhead{6.2} &
\colhead{6.2 ice} &
\colhead{11.3}
\\
\colhead{} & 
\colhead{[log $L/L_\sun$]} & 
\colhead{[log $L/L_\sun$]} & 
\colhead{[\um]} & 
\colhead{[\um]} & 
\colhead{[\um]} &
\colhead{[log $L/L_\sun$]} & 
\colhead{[log $L/L_\sun$]} & 
\colhead{[\um]} & 
\colhead{[\um]} & 
\colhead{[\um]}  
}
\startdata
IRAS 01355$-$1814 &    9.81  &   9.42 & 2.25    &        & 0.49    &    9.07  &   8.97  & 0.14    &        & 0.26    \\
IRAS 01418+1651   &    8.98  &   9.07 & 1.81    &        & 1.86    &    8.60  &   8.48  & 0.44    &        & 0.61    \\
IRAS 01562+2528   &    9.59  &   9.71 & 2.07    &        & 2.22    &    9.23  &   9.43  & 0.37    &        & 0.86    \\ 
IRAS 02524+2046   &    9.37  &   9.62 & 0.54    &        & 1.85    &    9.19  &   9.13  & 0.49    &        & 0.53    \\ 
IRAS 03521+0028   &    9.78  &   9.60 & 1.67    & 0.36   & 0.81    &    9.41  &   9.29  & 0.43    & 0.36   & 0.60    \\
IRAS 04121+0223   &    9.48  &   9.42 & 1.82    & 0.89   & 2.01    &    9.11  &   9.02  & 0.44    & 0.38   & 0.78    \\ 
IRAS 04454$-$4838 &   10.07  &   9.90 & 23.53   & 0.05   & 1.31    &    8.26  &   8.42  & 0.07    & 0.05   & 1.07    \\
IRAS 06487+2208   &    9.89  &   9.75 & 0.55    &        & 0.36    &    9.60  &   9.47  & 0.26    &        & 0.26    \\ 
IRAS 07163+0817   &    9.21  &   9.12 & 12.82   &        & 1.30    &    8.93  &   8.84  & 0.58    &        & 0.77    \\ 
IRAS 07572+0533   &     --   &   8.95 & --      &        & 0.04    &  $<8.95$ &   8.88  & $<0.10$ &        & 0.04    \\ 
IRAS 08201+2801   &   10.03  &   9.52 & 2.83    & 0.39   & 0.57    &    9.41  &   9.33  & 0.19    & 0.09   & 0.71    \\ 
IRAS 08449+2332   &    9.78  &   9.63 & 2.79    &        & 1.09    &    9.38  &   9.25  & 0.43    &        & 0.59    \\ 
IRAS 08474+1813   &    9.37  &   9.20 & 2.33    &        & 1.41    &    8.72  &   8.82  & 0.23    &        & 1.40    \\ 
IRAS 09039+0503   &    9.65  &   9.73 & 1.59    & 0.19   & 2.58    &    9.13  &   9.09  & 0.30    & 0.19   & 0.76    \\
IRAS 09539+0857   &   10.2   &  10.06 & 6.38    &        & 6.50    &    8.91  &   8.96  & 0.11    &        & 1.09    \\
IRAS 10035+2740   &    9.21  &   9.33 & 2.10    &        & 1.03    &    8.77  &   8.72  & 0.27    &        & 0.25    \\ 
IRAS 10039$-$3338 &   10.96  &  10.32 & 2.90    & 0.01   & 8.28    &    8.72  &   8.94  & 0.01    & 0.01   & 0.72    \\
IRAS 10173+0828   &    9.37  &   9.51 & 2.50    &        & 5.13    &    8.65  &   8.64  & 0.35    &        & 0.95    \\
IRAS 10339+1548   &    9.30  &   9.63 & 0.71    &        & 0.85    &    9.14  &   9.23  & 0.42    &        & 0.44    \\ 
IRAS 10378+1109   &    9.30  &   9.36 & 0.45    & 0.04   & 0.64    &    8.62  &   9.02  & 0.07    & 0.04   & 0.49    \\
IRAS 10485$-$1447 &    9.73  &   9.19 & 1.64    &        & 0.47    &    8.73  &   8.84  & 0.09    & 0.08   & 0.36    \\
IRAS 11028+3130   &    9.11  &   9.54 & 1.13    &        & 2.33    &    8.41  &   8.85  & 0.14    &        & 0.64    \\ 
IRAS 11180+1623   &    9.69  &   9.44 & 5.87    & 0.97   & 0.95    &    9.02  &   9.07  & 0.24    & 0.26   & 0.57    \\ 
IRAS 11524+1058   &    9.55  &   9.85 & 1.09    &        & 3.07    &    8.87  &   9.17  & 0.12    &        & 0.69    \\ 
IRAS 12018+1941   &    9.85  &   9.44 & 0.25    &        & 0.08    &    9.58  &   9.06  & 0.17    &        & 0.05    \\
IRAS 12032+1707   &   10.33  &  10.20 & 3.54    & 0.06   & 1.33    &    9.40  &   9.64  & 0.07    & 0.06   & 0.61    \\
IRAS 12112+0305   &    9.62  &   9.35 & 4.39    & 0.37   & 0.76    &    9.27  &   9.02  & 0.61    & 0.37   & 0.52    \\
IRAS 12540+5708   &    9.59  &   8.66 & 0.03    &        & 0.01    &    8.94  &   9.29  & 0.01    &        & 0.04    \\
IRAS 13218+0552   &      --  &   --   & --      & --     & --      &  $<9.79$ &   9.50  & $<0.01$ &        & $<0.03$ \\
IRAS 13428+5608   &    9.55  &   9.45 & 0.52    & 0.09   & 0.76    &    9.03  &   8.98  & 0.14    & 0.09   & 0.37    \\
IRAS 13451+1232   &    9.34  &   9.34 & 0.06    &        & 0.04    &    8.23  &   8.76  & 0.01    &        & 0.01    \\
IRAS 14059+2000   &    9.16  &   9.06 & 0.25    &        & 0.48    &    8.93  &   8.89  & 0.23    &        & 0.43    \\ 
IRAS 14070+0525   &   10.37  &  10.31 & 1.86    & 0.01   & 2.41    &    8.92  &   9.63  & 0.02    & 0.01   & 0.83    \\
IRAS 14553+1245   &    9.57  &   9.47 & 0.63    &        & 0.65    &    9.37  &   9.22  & 0.44    &        & 0.60    \\ 
IRAS 15327+2340   &   10.45  &  10.03 & 91.96   & 0.16   & 8.23    &    8.89  &   8.67  & 0.30    & 0.17   & 0.64    \\
IRAS 16090$-$0139 &   10.17  &   9.99 & 1.14    & 0.07   & 1.45    &    9.34  &   9.38  & 0.09    & 0.07   & 0.51    \\
IRAS 16255+2801   &    9.54  &   9.14 & 0.94    & 0.50   & 1.02    &    8.95  &   8.58  & 0.16    & 0.13   & 0.37    \\ 
IRAS 16300+1558   &   10.31  &   9.97 & 1.66    & 0.04   & 1.09    &    9.28  &   9.48  & 0.07    & 0.04   & 0.60    \\
IRAS 17207$-$0014 &   10.01  &   9.86 & 3.58    & 0.45   & 2.75    &    9.52  &   9.24  & 0.50    & 0.45   & 0.76    \\
IRAS 18368+3549   &    9.82  &   9.74 & 2.43    &        & 2.62    &    9.50  &   9.26  & 0.61    &        & 0.74    \\ 
IRAS 18588+3517   &    9.84  &   9.67 & 1.85    & 0.56   & 1.65    &    9.45  &   9.25  & 0.41    & 0.23   & 0.82    \\ 
IRAS 20100$-$4156 &    9.96  &   9.83 & 1.52    & 0.06   & 1.23    &    9.32  &   9.38  & 0.19    & 0.06   & 0.77    \\
IRAS 20286+1846   &    9.44  &   9.45 & 1.58    & 0.63   & 2.26    &    8.93  &   8.86  & 0.41    & 0.20   & 0.91    \\ 
IRAS 21077+3358   &    9.86  &   9.85 & 2.88    &        & 1.15    &    9.23  &   9.28  & 0.21    &        & 0.40    \\ 
IRAS 21272+2514   &    9.66  &   9.60 & 1.97    & 0.15   & 1.73    &    9.11  &   8.97  & 0.34    & 0.15   & 0.56    \\
IRAS 22055+3024   &    9.22  &   9.27 & 0.28    &        & 0.26    &    8.88  &   9.10  & 0.15    &        & 0.22    \\ 
IRAS 22116+0437   &   10.27  &   9.85 & 3.16    &        & 0.75    &    9.34  &   9.38  & 0.08    &        & 0.40    \\ 	
IRAS 22491$-$1808 &    9.47  &   9.33 & 1.38    & 0.45   & 0.98    &    9.07  &   8.94  & 0.43    & 0.45   & 0.57    \\
IRAS 23028+0725   &  \ldots  & \ldots & \ldots  & \ldots & \ldots  &  \ldots  &  \ldots & \ldots  & \ldots & \ldots  \\ 
IRAS 23233+0946   &    9.68  &   9.59 & 1.74    & 0.36   & 1.45    &    9.30  &   9.17  & 0.47    & 0.36   & 0.76    \\
IRAS 23365+3604   &    9.80  &   9.63 & 2.37    & 0.27   & 0.92    &    9.31  &   9.18  & 0.35    & 0.27   & 0.43    \\
\hline                                                   	                                   
IRAS 00163$-$1039 &    9.30  &   9.16 & 3.32    &        & 0.95    &    9.00  &   8.83  & 0.52    &        & 0.43    \\
IRAS 01572+0009   &    9.86  &   9.47 & 0.10    &	 & 0.04    &    9.60  &   9.55  & 0.07    &        & 0.06    \\
IRAS 05083+7936   &    9.93  &   9.95 & 4.45    &        & 2.10    &    9.68  &   9.57  & 0.62    &        & 0.66    \\
IRAS 06538+4628   &    8.75  &   8.77 & 1.00    &        & 0.60    &    8.51  &   8.50  & 0.45    &        & 0.36    \\
IRAS 08559+1053   &    9.89  &   9.86 & 0.31    & 0.19   & 0.55    &    9.63  &   9.52  & 0.21    & 0.19   & 0.29    \\
IRAS 09437+0317   &    9.10  &   9.16 & 1.47    &        & 1.99    &    8.89  &   8.82  & 0.62    &        & 0.77    \\
IRAS 10565+2448   &    9.90  &   9.80 & 1.53    &        & 1.43    &    9.55  &   9.31  & 0.51    &        & 0.51    \\
IRAS 11119+3257   &    9.92  &   9.68 & 0.04    & 0.05   & 0.04    &    9.93  &   8.36  & 0.06    & 0.05   & 0.003   \\
IRAS 13349+2438   &     --   &   --   & --      & --     & --      &  $<9.39$ & $<9.10$ & $<0.01$ &        & $<0.01$ \\
IRAS 15001+1433   &    9.86  &   9.86 & 0.30    & 0.14   & 0.43    &    9.54  &   9.44  & 0.16    & 0.14   & 0.20    \\
IRAS 15206+3342   &    9.81  &   9.75 & 0.38    &        & 0.41    &    9.56  &   9.47  & 0.25    &        & 0.27    \\
IRAS 20460+1925   &  \ldots  & \ldots & \ldots  & \ldots & \ldots  &  \ldots  & \ldots  & \ldots  & \ldots & \ldots  \\
IRAS 23007+0836   &    9.23  &   9.19 & 0.33    &	 & 0.34    &    8.99  &   8.89  & 0.19    &        & 0.20    \\
IRAS 23394$-$0353 &    9.26  &   9.16 & 2.24    &	 & 2.47    &    8.94  &   8.72  & 0.53    &        & 0.64    \\
IRAS 23498+2423   &    9.52  &   9.53 & 0.06    &	 & 0.13    &    9.25  &   9.05  & 0.04    &        & 0.06    \\
\enddata
\tablecomments{The ``ice'' 6.2~\um~PAH columns use a continuum that is corrected for water ice absorption (where present) at 6~\um. $-$ indicates that PAHFIT fit no significant flux for a particular dust component.}
\end{deluxetable*}

In addition to the atomic and simple molecular emission lines, we also observed multiple features attributed to polycyclic aromatic hydrocarbons (PAHs); the broad-line emission comes from vibrational modes of C-C and C-H bonds \citep{dra03}. PAH features are ubiquitous in the mid-IR emission of starburst galaxies and ULIRGs \citep{lut98,gen98,stu00,pee04,des07,ima07}, and dominate the low-resolution spectra of most galaxies in our sample. Multiple PAH features are seen for all OHMs, encompassing galaxies with very wide ranges in continuum shape and line emission. We detect strong PAH transitions centered at 6.2, 7.7, 8.6, 11.3, and 12.7~\um, several of which are also visible in the high-resolution spectra. Weaker emission features at 13.5, 14.2, 16.4, 17.1, and 17.4~\um~are also visible in many galaxies.

\begin{figure*}
\includegraphics[scale=1.0]{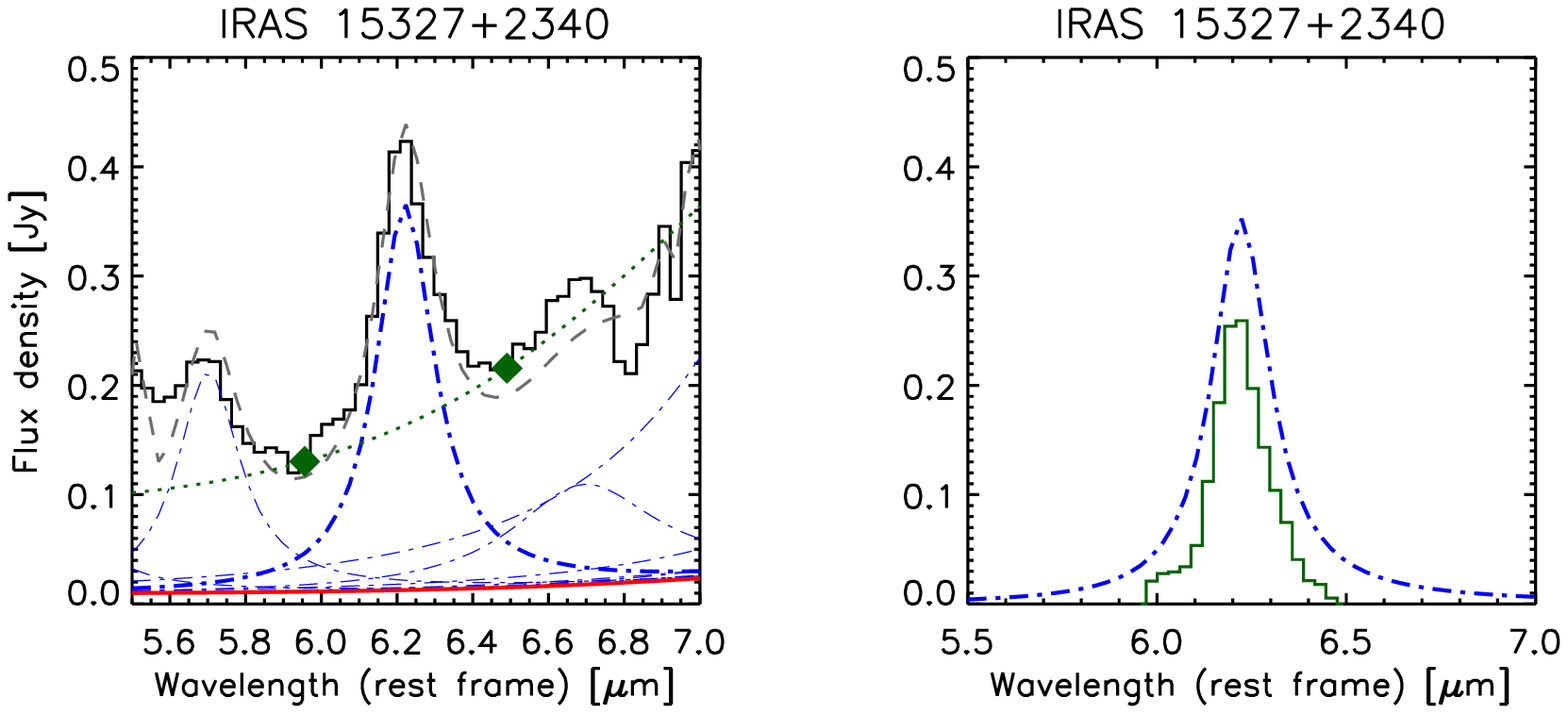}
\caption{Differences between two methods used to measure PAH fluxes in IRS spectra, illustrated on the spectrum of IRAS~15327+2340 (Arp~220). ({\it left}): The observed spectrum centered on the 6.2~\um~PAH feature is shown in black. The minimum at 5.9~\um~is the result of absorption by water ice. For the spline method, the PAH feature is defined as the integrated flux above a spline-interpolated continuum ({\it green dotted line}) between 5.95 and 6.5~\um. The dot-dashed blue lines represent individual Drude fit components from PAHFIT, while the red line is the continuum emission (blackbody dust + starlight). The dashed grey line shows PAHFIT's global fit to the data. ({\it right}): PAHFIT ({\it blue}) and spline-fit ({\it green}) results with the continuum subtracted; the feature as measured by PAHFIT has significantly more flux than with the spline continuum, typical of nearly all galaxies in our sample. \label{fig-pahcomp}}
\end{figure*}

We measured the PAH emission via two methods: the first defines a local continuum around the PAH feature using a spline fit, and then integrates the total flux after baseline subtraction. The default continuum pivots are located at 5.15, 5.55, 5.95, 6.55, and 7.10~\um~for the 6.2~\um~feature, and at 10.1, 10.9, 11.8, and 12.4~\um~for the 11.3~\um~feature. These are shifted slightly for each object to avoid both broad absorption features (including water ice and hydrocarbons) and narrow atomic emission lines. We quantify the emission from the two cleanest PAH features appearing in our spectra: the 6.2 and 11.3~\um~complexes (Table~\ref{tbl-pah}). 

A second method uses PAHFIT \citep{smi07}, a public IDL package, to fit global mid-IR spectral templates and simultaneously measure the relative effects of overlapping features. The routine decomposes the low-resolution IRS spectra into emission from stellar continuum, dust features (including PAHs), atomic and molecular lines, and blackbodies from thermally heated dust at a variety of temperatures; this is simultaneously fit with an extinction curve including silicate features at 9.7 and 18~\um. Due to the large number of parameters being fit, however, even strong features may overlap sufficiently to affect the accuracy of the fit \citep{spo02}; in addition, PAHFIT does not fit for several features commonly found in ULIRGs, such as absorption features from ice, hydrocarbons, and gas-phase molecules. 

PAHFIT fits the dust emission features (including PAHs) with a Drude profile, which has the form:

\begin{equation}
\label{eqn-drude}
I_\lambda[\lambda] = \frac{b \gamma^2}{(\lambda/\lambda_c - \lambda_c/\lambda)^2 + \gamma^2}, 
\end{equation}

\noindent where $\lambda_c$ is the central wavelength, $\gamma$ is the fractional FWHM, and $b$ is the (peak) central intensity. The Drude profile is typically broader than a Gaussian, with signifcant amounts of power in the extended wings. Several dust emission features ({\it eg}, the 7.7 and 12.7~\um~PAHs) require more than one component for a reasonable fit. PAHFIT returned positive detections for both the 6.2 and 11.3~\um~PAH features for nearly all galaxies; the fit for OHM IRAS~07572+0533 showed no emission at 6.2~\um, while the fits to OHM IRAS~13218+0552 and the non-masing galaxy IRAS~13349+2438 show no emission in either dust feature. 
 
We compared the flux measured in the 6.2 and 11.3~\um~PAH features from our baseline-subtracted spline fits to the PAHFIT values; results from PAHFIT are consistently higher than those from the spline fit, indicating significant mixing between the PAH emission and what was previously designated as ``continuum'' (Figure~\ref{fig-pahcomp}). Fluxes of the 6.2~\um~complex measured with PAHFIT are a factor of $\sim3-4$ greater than the spline-fit fluxes, while the 11.3~\um~feature is an average of $\sim2-3$ times larger. The relative strengths of the two features are consistent using both methods; the mean value of the (6.2~\um~PAH/11.3~\um~PAH) ratio is identical to within 15\% for OHMs.

\citet{gal08} also use both spline and multi-component profile fitting approaches to measure the PAH variations within galaxies; they show that both methods yield the same overall trends, although each have their own underlying biases depending on the property being measured. Given the large contributions of overlapping dust emission features to the 5--10~\um~spectrum, which are not possible to separate from the underlying blackbody emission using spline fits \citep{mar07}, we consider the PAHFIT results to be the more robust method. Measurements of PAH features in the literature, however, typically use a spline-fit method \citep[{\it eg},][]{bra06,des07,spo07,zak08}. In particular, comparisons of PAH data from the OHM galaxies to other samples \citep[{\it eg}, the ``fork diagram'' from][]{spo07} must use the same method to return physically meaningful results. While PAHFIT fluxes may thus better represent the absolute PAH luminosity, the spline-fit data are used when comparing the OHMs to objects from the literature (Table~\ref{tbl-pah}). 

%
Water ice absorption at 6~\um~can have significant effects on the measurement of the 6.2~\um~PAH equivalent width (EW); following the method of \citet{spo07}, we correct for this by substituting the continuum inferred while measuring the 9.7~\um~silicate strength for the measured 6.2~\um~continuum (see \S\ref{ssec-abs}). In total, 24 OHMs and three non-masing galaxies with SL data showed absorption strong enough to affect the measured EW; the spline fit with the new continuum decreased the EW for all objects except IRAS~11180+1623, for which the continuum levels as measured by the PAH fit and the silicate depth are nearly identical (within error). Since PAHFIT does not fit for ice absorption, we also calculate ice-corrected EW for objects showing absorption at 6~\um~by using the flux from PAHFIT and the inferred continuum from silicate measurements. 

The 11.3~\um~PAH is seated atop the edge of the deep silicate absorption at 9.7~\um; determining an extinction-corrected continuum level for EW measurements is thus also difficult. Spectral mapping of AGN galaxies with ISOCAM has shown that PAH emission can be spatially extended and suppressed near the nucleus \citep{lef01,dia10}; this means that the PAH emission in AGN may be largely unaffected by dust absorption. Starburst galaxies, however, can show strong PAH emission in both the 6.2 and 11.3~\um~bands in the nuclear regions where the silicate optical depth is at its highest \citep{gal08}, and is likely to affect continuum levels for PAH features. The measured 11.3~\um~PAH data using the spline-fit method are therefore likely to underestimate the luminosities.


\subsection{Absorption features}\label{ssec-abs}

\begin{figure*}
\includegraphics[scale=1.0]{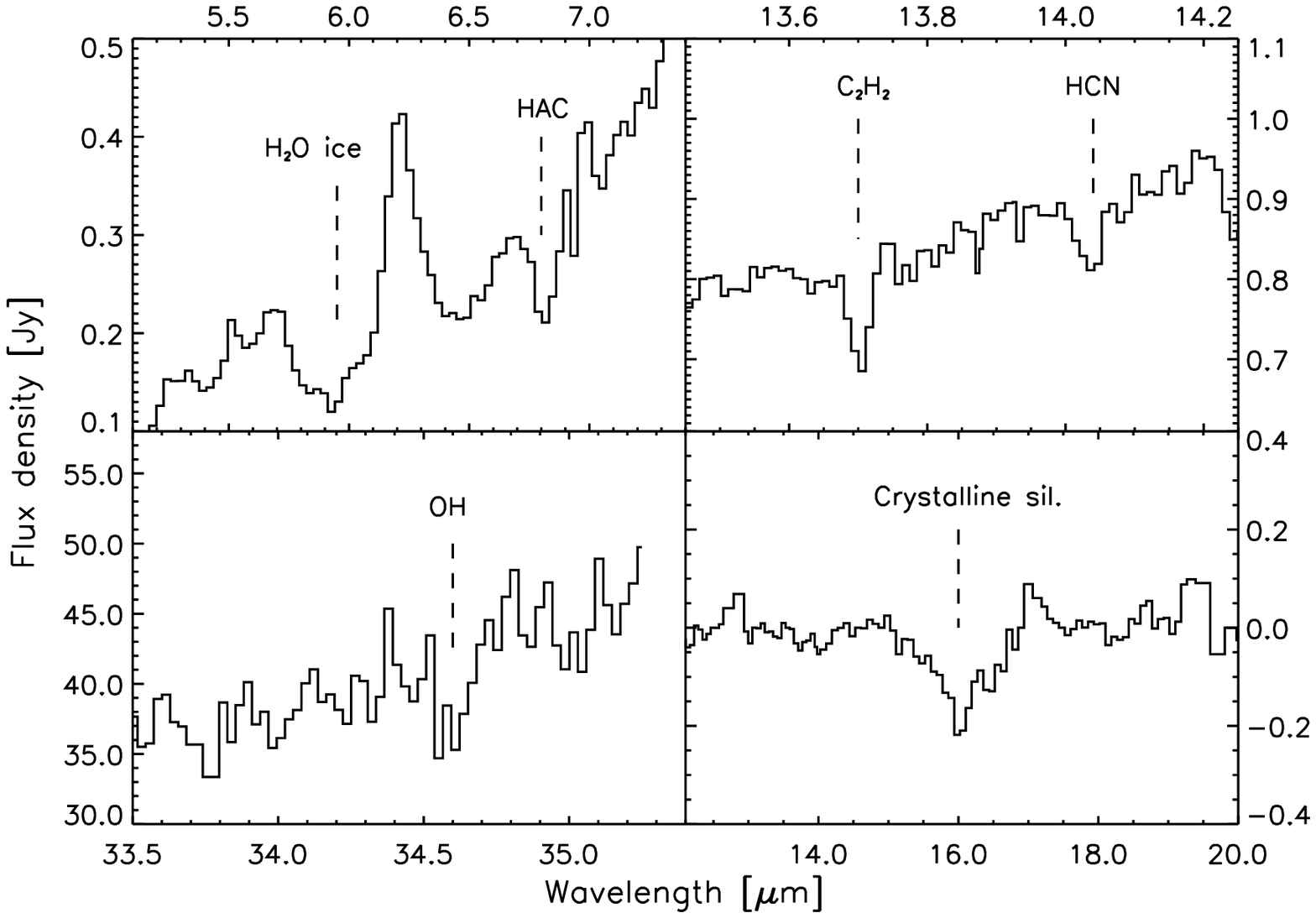}
\caption{Examples of mid-IR absorption features for the OHM IRAS~15327+2340 (Arp~220). {\it Top left}: The 6.0~\um~\water~ice and 6.85~\um~HAC absorption features. {\it Top right}: Gas-phase \acet~13.7~\um~and HCN 14.0~\um~absorption. {\it Bottom left}: Gas-phase OH 34.6~\um~absorption. {\it Bottom right}: Residual optical depth of the 16~\um~crystalline silicate feature. Measurements for all labeled features are in Tables~\ref{tbl-solidabs} and \ref{tbl-gasabs}. \label{fig-abs_example}}
\end{figure*}

\subsubsection{Silicates}\label{sssec-sil}

\begin{deluxetable*}{lcllllcccc}
\tabletypesize{\scriptsize}
\tablecaption{Solid-phase absorption features\label{tbl-solidabs}}
\tablewidth{0pt}
\tablehead{
\colhead{Object} & 
\colhead{\underline{6.0~\um~H$_2$O ice}} & 
\multicolumn{2}{c}{\underline{6.85~\um~HAC}} &
\multicolumn{2}{c}{\underline{7.25~\um~HAC}} &
\colhead{\underline{9.7~\um}} & 
\colhead{\underline{18~\um}} &
\colhead{\underline{16~\um}} & 
\colhead{\underline{23~\um}}
\\
\colhead{} & 
\colhead{$\tau$} & 
\colhead{$\tau$} & 
\colhead{flux} &
\colhead{$\tau$} & 
\colhead{flux} &
\colhead{$S_{sil}$} &
\colhead{$S_{sil}$} &
\colhead{$S_{sil}^{resid}$} &
\colhead{$S_{sil}^{resid}$}
}
\startdata
IRAS 01355$-$1814 &        &        &          &        &          & $-$2.4 & $-$0.9  & $-$0.3  & $-$0.2 \\ 
IRAS 01418+1651   &        & 0.49   & $-$8.3   &        &          & $-$1.3 & $-$0.4  &         &        \\ 
IRAS 01562+2528   &        &        &          &        &          & $-$0.7 & $-$0.3  &         &        \\ 
IRAS 02524+2046   &        & 0.20   & $-$0.4   &        &          & $-$0.9 & $-$0.3  &         &        \\ 
IRAS 03521+0028   & 0.42   & 0.13   & $-$0.4   &        &          & $-$1.4 & $-$0.2  & $-$0.1  & $-$0.1 \\ 
IRAS 04121+0223   & 1.61   &        &          &        &          & $-$1.0 & $-$0.2  &         &        \\ 
IRAS 04454$-$4838 & 0.42   & 0.35   & $-$9.0   & 0.10   & $-$1.5   & $-$3.7 & $-$1.0  & $-$0.3  & $-$0.2 \\ 
IRAS 06487+2208   &        & 0.23   & $-$4.6   &        &          & $-$1.2 & $-$0.3  &         &        \\ 
IRAS 07163+0817   &        &        &          &        &          & $-$1.2 & $-$0.1  &         &        \\ 
IRAS 07572+0533   &        &        &          &        &          & $-$0.6 & $-$0.3  &         &        \\ 
IRAS 08201+2801   & 1.06   & 0.37   & $-$4.0   & 0.22   & $-$1.5   & $-$2.2 & $-$0.6  & $-$0.2  &$-$0.1  \\ 
IRAS 08449+2332   &        &        &          &        &          & $-$1.2 & $-$0.5  & $-$0.1  &$-$0.1  \\ 
IRAS 08474+1813   &        & 0.36   & $-$0.2   &        &          & $-$1.9 & $-$1.2  &         &        \\ 
IRAS 09039+0503   & 0.98   & 0.15   & $-$0.5   &        &          & $-$2.0 & $-$0.6  &         &        \\ 
IRAS 09539+0857   &        & 0.24   & $-$2.5   &        &          & $-$3.1 & $-$1.2  & $-$0.4  &$-$0.2  \\ 
IRAS 10035+2740   &        &        &          &        &          & $-$1.5 & $-$0.8  &         &        \\ 
IRAS 10039$-$3338 & 0.23   & 0.23   & $-$166.3 &        &          & $-$3.1 & $-$1.0  & $-$0.4  &$-$0.3  \\ 
IRAS 10173+0828   &        &        &          &        &          & $-$1.9 & $-$0.8  & $-$0.3  &$-$0.2  \\ 
IRAS 10339+1548   &        &        &          &        &          & $-$1.1 & $-$0.05 &         &        \\ 
IRAS 10378+1109   & 0.72   & 0.18   & $-$0.6   &        &          & $-$2.0 & $-$0.3  &         &        \\ 
IRAS 10485$-$1447 &        & 0.25   & $-$0.4   &        &          & $-$2.9 & $-$0.9  &         &        \\ 
IRAS 11028+3130   &        &        &          &        &          & $-$2.6 & $-$1.0  &         &        \\ 
IRAS 11180+1623   & 0.54   &        &          &        &          & $-$1.7 & $-$0.5  &         &        \\ 
IRAS 11524+1058   &        & 0.27   & $-$1.5   & 0.2:   & $-$0.3:  & $-$1.5 & $-$0.8  &         &        \\ 
IRAS 12018+1941   &        & 0.16   & $-$2.6   &        &          & $-$1.4 & $-$0.4  &         &        \\ 
IRAS 12032+1707   & 0.71   & 0.55   & $-$6.6   & 0.30   & $-$3.9   & $-$2.7 & $-$0.8  &         &        \\ 
IRAS 12112+0305   & 0.59   & 0.41   & $-$3.5   &        &          & $-$1.8 & $-$0.3  & $-$0.2  &$-$0.1  \\ 
IRAS 12540+5708   &        &        &          &        &          & $-$0.7 & $-$0.2  &         &        \\ 
IRAS 13218+0552   &        &        &          &        &          & $-$0.5 & $-$0.4  &         &        \\ 
IRAS 13428+5608   & 0.50   & 0.40   & $-$28.5  &        &          & $-$2.0 & $-$0.5  &         &        \\ 
IRAS 13451+1232   &        &        &          &        &          & $-$0.5 & $-$0.1  &         &        \\ 
IRAS 14059+2000   &        &        &          &        &          & $-$0.8 & $-$0.1  &         &        \\ 
IRAS 14070+0525   & 0.90   & 0.24   & $-$1.8   & 0.15   & $-$1.8   & $-$2.7 & $-$0.9  &         &        \\ 
IRAS 14553+1245   &        &        &          &        &          & $-$1.3 & $-$0.5  &         &        \\ 
IRAS 15327+2340   & 0.68   & 0.35   & $-$50.1  &        &          & $-$3.1 & $-$0.4  & $-$0.2  &$-$0.1  \\ 
IRAS 16090$-$0139 & 0.56   & 0.45   & $-$10.5  & 0.24   & $-$5.6   & $-$2.4 & $-$0.6  &         &        \\ 
IRAS 16255+2801   & 0.54   &        &          &        &          & $-$2.2 & $-$0.6  &         &        \\ 
IRAS 16300+1558   & 0.61   & 0.38   & $-$2.2   & 0.21   & $-$1.1   & $-$2.7 & $-$0.7  & $-$0.3  &$-$0.1  \\ 
IRAS 17207$-$0014 & 0.31   & 0.23   & $-$17.4  &        &          & $-$1.9 & $-$0.6  & $-$0.2  &$-$0.1  \\ 
IRAS 18368+3549   &        &        &          &        &          & $-$1.8 & $-$0.2  & $-$0.3  &$-$0.2  \\ 
IRAS 18588+3517   & 0.72   &        &          &        &          & $-$2.2 & $-$0.6  & $-$0.3  & -- \\ 
IRAS 20100$-$4156 & 1.45   & 0.23   & $-$2.9   & 0.23   & $-$4.5   & $-$2.4 & $-$0.7  & $-$0.2  &$-$0.1  \\ 
IRAS 20286+1846   & 1.08   &        &          &        &          & $-$1.6 & $-$0.6  &         &        \\ 
IRAS 21077+3358   &        &        &          &        &          & $-$1.9 & $-$0.7  &         &        \\ 
IRAS 21272+2514   & 1.66   & 0.21   & $-$0.7   &        &          & $-$2.8 & $-$0.7  &         &        \\ 
IRAS 22055+3024   &        &        &          &        &          & $-$1.3 & $-$0.3  &         &        \\ 
IRAS 22116+0437   &        & 0.04:  & $-$0.1:  &        &          & $-$2.6 & $-$0.9  & $-$0.2  &$-$0.2  \\ 
IRAS 22491$-$1808 & 0.43   & 0.19   & $-$0.2   &        &          & $-$1.5 & $-$0.5  & $-$0.2  & -- \\ 
IRAS 23028+0725   & \ldots & \ldots & \ldots   & \ldots & \ldots   & \ldots & $-$0.6  &         &        \\ 
IRAS 23233+0946   & 0.44   &        &          &        &          & $-$1.9 & $-$0.4  &         &        \\ 
IRAS 23365+3604   & 0.66   &        &          &        &          & $-$2.0 & $-$0.5  &         &        \\ 
\hline                                                                                     
IRAS 00163$-$1039 &        &        &          &        &          & $-$0.5 & $-$0.1  &         &        \\ 
IRAS 01572+0009   &        &        &          &        &          & $-$0.2 & $-$0.2  &         &        \\ 
IRAS 05083+7936   &        &        &          &        &          & $-$1.1 & $-$0.3  &         &        \\ 
IRAS 06538+4628   &        &        &          &        &          & $-$0.5 & $-$0.2  &         &        \\ 
IRAS 08559+1053   & 0.18   &        &          &        &          & $-$0.6 & $-$0.2  &         &        \\ 
IRAS 09437+0317   &        &        &          &        &          & $-$1.1 & $-$0.3  &         &        \\ 
IRAS 10565+2448   &        &        &          &        &          & $-$1.2 & $-$0.3  &         &        \\ 
IRAS 11119+3257   & 0.19   &        &          &        &          & $-$0.7 & $-$0.3  & $-$0.2  & -- \\ 
IRAS 13349+2438   &        &        &          &        &          &    0.1 &    0.07 &          &       \\ 
IRAS 15001+1433   & 0.30   &        &          &        &          & $-$0.9 & $-$0.4  &          &       \\ 
IRAS 15206+3342   &        &        &          &        &          & $-$0.4 & $-$0.2  &          &       \\ 
IRAS 20460+1925   & \ldots & \ldots & \ldots   & \ldots &  \ldots  & \ldots & $-$0.4  &          &       \\ 
IRAS 23007+0836   &        &        &          &        &          & $-$0.3 & $-$0.1  &          &       \\ 
IRAS 23394$-$0353 &        &        &          &        &          & $-$0.7 & $-$0.3  &          &       \\ 
IRAS 23498+2423   &        &        &          &        &          & $-$0.6 & $-$0.3  &          &       \\ 
\enddata
\tablecomments{Silicate strength is defined in Eqn.~\ref{eqn-silstrength}; the 9.7 and 18~\um~features are depths for amorphous silicates, while the 16 and 23~\um~crystalline features are residual depths measured after the 18~\um~feature was subtracted. Fluxes are given in $10^{-21}$~W~cm$^{-2}$; objects marked with a : represent uncertain detections.}
\end{deluxetable*}

The LR spectra show near-ubiquitous absorption from amorphous silicate dust, with a strong feature caused by a Si-O stretching mode near 9.7~\um~and a weaker feature caused by an Si-O-Si bending mode near 18~\um~\citep{kna73}. The presence of dust is unsurprising, as the characteristic extreme IR luminosities of ULIRGs are caused by large amounts of heated dust being thermally re-radiated. \cite{hao07} found that ULIRGs nearly uniformly show absorption in the two silicate features, in contrast to QSOs and some Seyfert galaxies which typically show the feature in emission \citep{sie05,hao05,stu05,sch08}. 

We measure the strength of the silicate absorption at both 9.7 and 18~\um~using the method of \citet{spo07}:

\begin{equation}
\label{eqn-silstrength}
S_{\lambda} = ln\left(\frac{f_{\lambda}}{f_{cont}}\right),
\end{equation}

\noindent where $f_{\lambda}$ is the measured flux and $f_{cont}$ the interpolated continuum at the feature extremum. More negative values of $S_{sil}$ represent deeper absorption. The expected continuum is calculated using a combination of spline and power-law fits, depending on the strength of the PAH and water ice features in the spectrum \citep{spo07}. 

All OHMs and $\sim95\%$ of the non-masing galaxies showed absorption at both 9.7 and 18~\um~(Table~\ref{tbl-solidabs}); the average depth for the OHMs is $S_{9.7}=-1.8\pm0.8$, while the average depth of the non-masing galaxies is $S_{9.7}=-0.6\pm0.4$. The deepest absorption is in the OHM IRAS~04454-0838 ($S_{9.7}=-3.7$), while only one object (the non-masing galaxy IRAS~13349+2438) shows emission in both amorphous silicate features. 

In addition to the amorphous silicate, we detect weaker features from crystalline silicate absorption in 19 OHMs, including bands at 11, 16, 19, 23, and 28~\um~(see Figure~\ref{fig-abs_example} for an example). We use the method of \citet{spo06} to subtract off both the dust continuum and amorphous component to measure the residual optical depth at 16 and 23~\um, typically the strongest crystalline features (Table~\ref{tbl-solidabs}). The deepest $S_{16}$ occurs for IRAS~10039$-$3338, at $-0.4$; however, the S/N ratio means we are only sensitive to an absorption limit of $S_{16}\simeq-0.1$ and $S_{18}\simeq-0.05$. Only one detection of crystalline silicates is made in a non-masing galaxy, in IRAS~11119+3257. Since all detections of crystalline silicates in OHMs have $S_{9.7}<-1.4$, the lower total dust column in non-masing galaxies is a likely contributor to the detection rate. 

\subsubsection{Aliphatic hydrocarbons}\label{sssec-hac}

Absorption bands arising from hydrogenated amorphous carbon grains (HACs) can also be significant contributors to diffuse dust in the galactic ISM \citep{chi00}. ISO \citep{spo01,spo02} and \textit{Spitzer} \citep{dar07b,dar07c} observations have identified HAC absorption features due to bending modes at 6.85~\um~(CH$_2$/CH$_3$) and 7.25~\um~(CH$_3$) in ULIRGs. These aliphatic features represent a counterpart to the aromatic hydrocarbons responsible for PAH emission and are an abundant component of the ISM in luminous galaxies. 

We detect absorption from the 6.85~\um~HAC transition in 27/51 galaxies in the OHM sample (Figure~\ref{fig-abs_example}), with zero detections in the non-masing sample. The accopmanying 7.25~\um~feature is detected in eight of the galaxies in which the 6.85~\um~feature is seen. The strength of the HAC is measured using a spline fit with pivots at 5.2, 5.6, 7.8, 14.0, and 26.0~\um~to determine the local continuum in the 6--8~\um~region \citep{spo07} and integrate the total flux within the absorption feature (Table~\ref{tbl-solidabs}). The average optical depth of the 6.85~\um~feature is $\tau_{6.85}=0.23\pm0.11$, and the average depth of the 7.25~\um~feature is $\tau_{7.25}=0.20\pm0.06$. Many of the galaxies with no detection of HACs, however, have limits from noise that are consistent with the absorption depths measured in brighter galaxies. 

\subsubsection{Ices}\label{sssec-ices}

Absorption from ices in a variety of molecular species (including \water, CO, \cotwo, and \meth) has been detected in spectroscopy of IR-bright galaxies \citep{spo00,spo01,stu00}. The band from water ice absorption stretching from 6--8~\um~is prominent (Figure~\ref{fig-abs_example}) and was detected in $\sim10$\% of a sample of bright galaxies using ISO \citep{spo02} and the IRS \citep{arm04,spo05,arm07}. 

We detected water ice absorption at 6~\um~in 24 OHMs and three non-masing galaxies. We use the spline continuum from fitting the 9.7~\um~silicate feature as the local 5.5--7~\um~continuum in order to obtain an optical depth spectrum for the 6~\um~absorption complex. The resulting water ice optical depths are tabulated in Table~\ref{tbl-solidabs}. We note that contamination by 6.2~\um~PAH emission and absorption by other species than water ice \citep{spo05} may add confusion in properly measuring optical depths. 

\begin{deluxetable}{lllllllll}
\tabletypesize{\scriptsize}
\tablecaption{Gas-phase absorption features\label{tbl-gasabs}}
\tablewidth{0pt}
\tablehead{
\colhead{Object} & 
\multicolumn{2}{c}{\underline{13.7~\um~\acet}} &
\multicolumn{2}{c}{\underline{14.02~\um~HCN}} &
\multicolumn{2}{c}{\underline{15.0~\um~\cotwo}} &
\\
\colhead{} & 
\colhead{$f_{norm}$} & 
\colhead{flux} &
\colhead{$f_{norm}$} & 
\colhead{flux} &
\colhead{$f_{norm}$} & 
\colhead{flux} &
}
\startdata
IRAS 10039$-$3338 & 0.93 & $-$0.9  & 0.86 & $-$4.0  &      &        \\ 
IRAS 12018+1941   & 0.95 & $-$0.7  & 0.94 & $-$0.8  &      &        \\ 
IRAS 12540+5708   & 0.95 & $-$9.5  &      &         &      &        \\ 
IRAS 13218+0552   & 0.82 & $-$1.1  &      &         &      &        \\ 
IRAS 13428+5608   & 0.93 & $-$0.8  &      &         &      &        \\ 
IRAS 14070+0525   & 0.79 & $-$0.4  &      &         &      &        \\ 
IRAS 15327+2340   & 0.84 & $-$5.8  & 0.90 & $-$5.6  & 0.94 & $-$2.0 \\ 
IRAS 16090$-$0139 & 0.86 & $-$1.1  &      &         &      &        \\ 
IRAS 17207$-$0014 & 0.93 & $-$0.7  &      &         &      &        \\ 
IRAS 20100$-$4156 & 0.82 & $-$1.4  & 0.84 & $-$1.1  &      &        \\ 
IRAS 22491$-$1808 & 0.92 & $-$0.7  &      &         &      &        \\ 
\enddata
\tablecomments{$f_{norm}$ gives the peak depth of absorption features plotted in normalized flux units. Fluxes are measured in $10^{-21}$~W~cm$^{-2}$.}
\end{deluxetable}

\subsubsection{\acet, \hcn, and \cotwo}\label{sssec-molgas}

Previous mid-IR surveys have also identified bands of molecular gas absorption in ULIRGs \citep{spo06,arm07,lah07}, including the vibration-rotation bands of acetylene (\acet; 13.7~\um), hydrogen cyanide (HCN; 14.02~\um), and carbon dioxide (\cotwo; 15.0~\um). \citet{lah07} reported the detection of both \acet~and HCN in fifteen (U)LIRG nuclei, with detections of \cotwo~in four objects. Eight of the objects in the Lahuis sample are OHMs in our sample; we confirm detections of \acet~in all eight galaxies, in addition to the OHMs IRAS~10039-3338 and IRAS~12018+1941 (Figure~\ref{fig-abs_example}). HCN is detected in only 4/10 archival galaxies, meaning that we cannot confirm the HCN detection of four galaxies; since the optical depth of HCN is typically much weaker than that of \acet, however, it is possible that our lower detection rate is a result of improved S/N in their reduction process. We also confirm the detection of \cotwo~in IRAS~15327+2340 (Arp~220). No galaxies in the non-masing control sample showed absorption in any molecular band, nor did any of the OHMs observed in our dedicated program. 

Since these gas-phase absorption features are actually a blend of multiple absorption lines, the peak optical depth measured is a function of the velocity resolution of the spectrograph. We therefore report the integrated flux and the peak depth in normalized flux units \citep[$f_{norm}$, where the spectrum has been divided by the adopted continuum;][]{spo04} in Table~\ref{tbl-gasabs}. 

	\cite{lah07} model abundances for ULIRGs with detections of \acet, HCN, and \cotwo, and suggest that they are associated with a phase of deeply embedded star formation, excluding the possibility of the features arising from an X-ray dominated region (XDR) powered by AGN. \cite{dar07} has also shown that OHMs have the highest mean molecular gas densities among starburst galaxies (traced by the $J=1\rightarrow0$ rotational HCN transition) and also possess high dense molecular gas fractions, comprising a distinct population in the IR-CO relation. The results of \cite{lah07} show that 9/15 ULIRGs with absorption in both \acet~and HCN are known OHMs in our {\it Spitzer} sample; this dense gas fraction ($\sim50\%$) is nearly identical to the observed OHM fraction in starbursts with dense (ULIRGs with $L_{HCN}/L_{CO} > 0.07$) fractions of molecular gas \citep{gao04,dar07,baa08}. 

\begin{figure}
\includegraphics[scale=0.5]{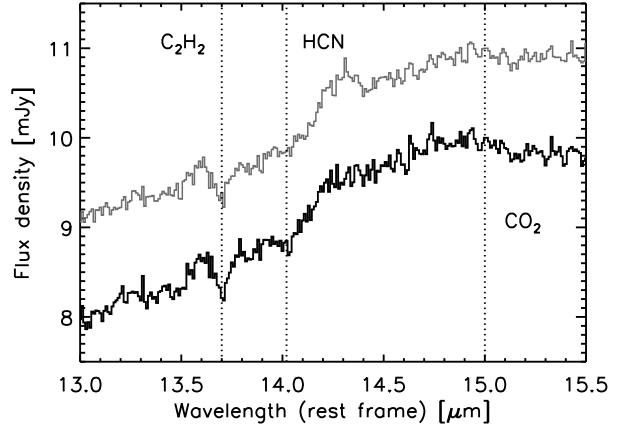}
\caption{Medianed HR spectra for both OHMs ({\it black}) and non-masing ({\it grey}) galaxies, sampled at intervals of $0.01$~\um~and normalized in flux at 15~\um. PAH emission is visible in bands centered at 13.6 and 14.2~\um. The dotted lines mark locations of gas-phase absorption in \acet~(13.7~\um), \hcn~(14.02~\um), and \cotwo~(15.0~\um). The spectra are vertically offset to highlight the differences between the samples. \label{fig-hac_avg}}
\end{figure}

	Given that the S/N ratio for the OHM and non-masing galaxies are of comparable magnitude, the lack of detection of any gas-phase species in the non-masing galaxies is a striking difference compared to the OHMs. Figure~\ref{fig-hac_avg} shows the median stack of both samples near the regions of gas-phase absorption; while the \acet~feature at 13.7~\um~can be clearly seen in the median OHM spectrum, neither the \hcn~nor the \cotwo~transition is prominent. Since the data have been median stacked (as opposed to a mean, which can be dominated by a few deep absorbers), this suggests low-levels of \acet~present in a significant fraction of the OHM host galaxies. No molecular absorption appears in the medianed spectrum for the non-masing galaxies. 

	While a connection between OHMs and dense molecular gas is known to exist, the lack of detected molecular absorption in the mid-IR for the majority of OHMs is not entirely unexpected. OHMs occur in merging galaxies where different populations of gas may be kinematically and thermally distinct, yet are observed as a single unresolved region within the \textit{Spitzer} beam. \cite{lah07} suggest that high abundances of warm, dense gas are associated with deeply-embedded star formation, where \HII~regions are prevented from expanding by large pressure gradients and extend the lifetime of the star formation process. 

	\cite{baa08} interpret dense gas abundances as excluding very hard radiation fields (such as those found in XDRs) that dissociate the molecules; they suggest that the molecular emission arises from PDRs surrounding \HII~regions. Although few OHMs show absorption from dense molecular gas, the OHM sample also has few identified AGN or XDRs (only 4/51 OHMs show \neV~at 14~\um). The connection between dense molecular gas and the presence of an AGN is thus unclear based on this data alone. 
	
\subsubsection{Gas-phase OH}\label{sssec-ohgas}

\begin{deluxetable*}{lcccrrrc}
\tabletypesize{\scriptsize}
\tablecaption{Properties of OH gas-phase absorption\label{tbl-ohabs}}
\tablewidth{0pt}
\tablehead{
\colhead{OHM} &
\colhead{$f_{OH}$} &
\colhead{$\tau_{peak}$} &
\colhead{EW} &
\colhead{$N_{OH}$} &
\colhead{$\gamma_{abs}$} &
\colhead{$\gamma_{OHM}$} &
\colhead{$\phi_{pump}$}
\\
\colhead{} &
\colhead{[10$^{-21}$ W cm$^{-2}$]} &
\colhead{} &
\colhead{[$10^{-3}$~\um]} &
\colhead{[cm$^{-2}$]} &
\colhead{[ph/s]} &
\colhead{[ph/s]} &
\colhead{\%}
}
\startdata
IRAS 01418+1651 (III Zw 35) &   $-$6.1 & 0.15 &  5.5 & $1.1\times10^{17}$ & $1.7\times10^{54}$ & $1.8\times10^{53}$ &  10  \\
IRAS 13428+5608 (Mrk 273)   &  $-$28.5 & 0.14 & 10.2 & $2.1\times10^{17}$ & $1.7\times10^{55}$ & $1.4\times10^{53}$ &  0.8 \\
IRAS 15327+2340 (Arp 220)   &   $-$172 & 0.21 & 15.0 & $3.0\times10^{17}$ & $2.3\times10^{55}$ & $1.6\times10^{53}$ &  0.7 \\
\enddata
\tablecomments{The pumping efficiency ($\phi_{pump}=\gamma_{OHM}/\gamma_{abs}\times100$) assumes all pumping comes from the 34.6~\um~transition.}
\end{deluxetable*}

For three OHMs, we report detection of the $^2\Pi_{1/2}$~J=$5/2\rightarrow$~$^2\Pi_{3/2}$~J=3/2 OH absorption doublet near 34.616~\um: III~Zw~35 (IRAS~01418+1651), Mrk~273 (IRAS~13428+5608), and Arp~220 (IRAS~15327+2340). This feature is generally difficult to detect since it lies near the noisy, far-red edge of the LH module; for all objects with $z>0.08$, it is redshifted out of the IRS range. OH absorption at 34.6~\um~in Arp~220 was first reported by \citet{ski97} using ISO and confirmed with the IRS by \citet{far07}, who incorrectly identified it as the OH$^-$ ion. All OH absorption features are well fit with a single Gaussian, since the separation between the doublet features ($\Delta\lambda\simeq0.02~\mu$m) is comparable to the resolution element in the LH module. No detection of OH absorption was made for any of the non-masing galaxies. 

	Assuming the OH transitions are optically thin, we can use the EW to derive a column density for the OH ground state, which is likely to be a good proxy for the total column at typical molecular cloud densities \citep{bra99}:

\begin{equation}
\label{eqn-ohcolumn}
N_l = \frac{EW}{A_{ul}} \frac{8 \pi c}{\lambda^4} \frac{g_l}{g_u}.
\end{equation}

\noindent OH column densities for all galaxies are quite similar, lying between $(1-3)\times 10^{17}$~cm$^{-2}$ (Table~\ref{tbl-ohabs}). The measured $N_{OH}$ from the IRS data for Arp~220 also agrees within a factor of 2 of the column measured with ISO \citep{ski97}. Limits for galaxies in which the 34.6~\um~OH feature is not detected are of order $N_{OH}\lesssim1\times10^{17}$~cm$^{-2}$. 

	We compare the $N_{OH}$ derived from the rotational 34.6~\um~transitions to the OH column density measured in galaxies who show the hyperfine 1667~MHz feature in absorption. The majority of such galaxies are ULIRGs of comparable luminosity to the galaxies in our non-masing sample. Measurements from ten OH absorbers \citep{baa92,dar07} give $N_{OH}=T_{ex}(1.8\pm1.9)\times10^{15}$~cm$^{-2}$, where $T_{ex}$ is the OH excitation temperature in K. If the dust and gas are well-mixed, then the temperature of the dust ($\sim50-100$~K) can be used as a proxy for $T_{ex}$. This gives OH column densities for both OHMs (34.6~\um) and non-masing galaxies (1667~MHz) with comparable values of $N_{OH}\simeq10^{17}$~cm$^{-2}$. If so, then this addresses one of the crucial differences between OHMs and non-masing ULIRGs - namely, that differences in the abundance of masing molecules are {\it not} a key factor for triggering an OHM. 
	
\begin{figure*}
\includegraphics[scale=0.95]{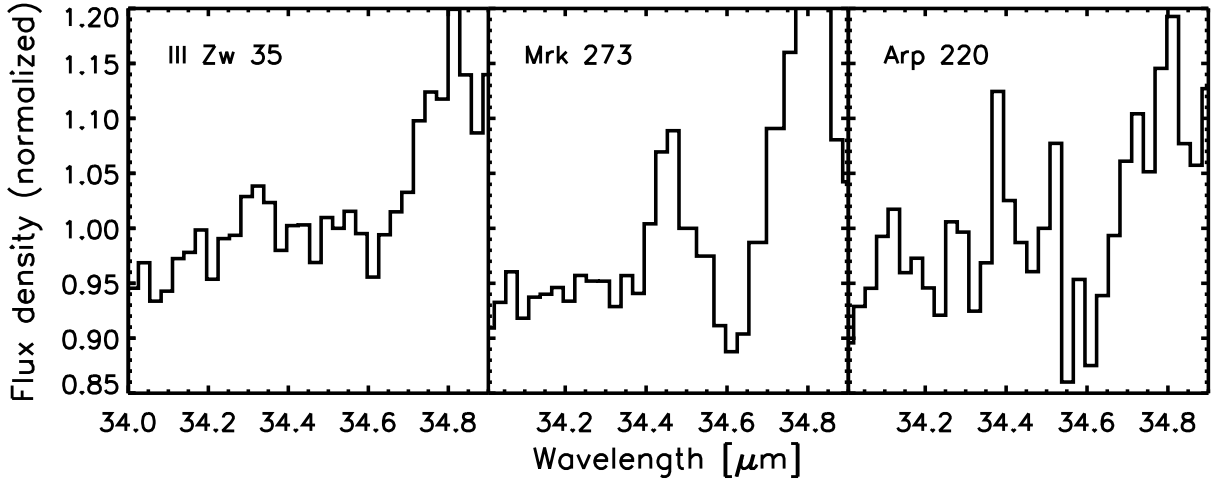}
\caption{The 34.6~\um~OH absorption feature in III~Zw~35, Mrk~273, and Arp~220. All spectra are been normalized in flux near 34.5~\um. \label{fig-ohabsplot}}
\end{figure*}

	The amount of OH available in the galaxy can also test models of the OHM pumping mechanism. \citet{ski97} computed the photon flux ($\gamma_{abs}=L_{abs}^{OH}/h\nu_{OH}$) absorbed in the 34.6~\um~transition from the OHM Arp~220. They found that $\gamma_{abs}$ is roughly 1\% of the photon flux in the OHM ($\gamma_{OHM}=L_{OHM}/h\nu_{OHM}$). If the 18-cm and mid-IR pumping photons lie along the same line of sight, this means that pumping photons from the 34.6~\um~transition alone can power the OHM (given an efficiency of $\sim1\%$ or higher). While radiative transfer models from \citet{loc08} suggest that the 53~\um~OH transition likely contributes more pumping photons than the 34.6~\um~line, the energetics are consistent with the basic acccepted mid-IR pumping model. 
	
	For the galaxies in our sample with OH absorption, Arp~220 and Mrk~273 would require pumping efficiencies on the order of 1\% to power the OHMs from the 34.6~\um~transition alone. III~Zw~35 shows the weakest OH absorption among our detections and would require an efficiency of $\phi_{pump}\simeq10\%$. 


\section{Conclusions}\label{sec-conclusion}

We present mid-infrared spectra and photometry for 51 OH megamasers taken with the IRS on \textit{Spitzer}, along with 15 galaxies confirmed to have {\bf no} megamaser emission above $L_{OH}=10^{2.3} L_\sun$. All objects in both samples have full coverage in both the low- and high-resolution IRS modules. We measure both emission (PAH, \htwo, and fine-structure atomic transitions) and absorption (silicates, hydrogenated amorphous carbon grains, and molecular bands) features, with full spectra, line fluxes, equivalent widths, and absorption depths presented for each object. 

The majority of the galaxies closely resemble standard mid-IR ULIRG templates, with the low-resolution emission dominated by moderate-to-deep amorphous silicate absorption at 9.7 and 18~\um~and PAH features at 6.2, 7.7, 8.6, 11.3 and 12.7~\um. The OHMs (on average) show deeper silicate absorption and steeper continuum slopes than the non-masing galaxies. Crystalline silicate absorption is detected in roughly a third of OHMs, but in only 1 out of 15 non-masing galaxies. OHMs are also the only galaxies in our sample to show absorption from hydrogenated amorphous carbon (HAC), gas-phase \hcn, \acet, and \cotwo; however, higher average noise in the non-masing spectra mean that features at similar absorption depths could be obscured due to lack of sensitivity. High-resolution spectra show emission from \neII~and \neIII~in almost all galaxies, with emission from \sIII, \sIV, and \oIV~commonly detected. The high-ionization \neV~line (a clear tracer of AGN) is detected in $<10\%$ of OHMs and in 53\% of the non-masing galaxies. Almost all galaxies in both samples also show emission in multiple \htwo~rotational transitions. 

We also measure the 34.6~\um~OH transition in three OHMs. OH column densities derived from the mid-IR OH transition are of the same order of magnitude as the column densities derived from the 1667~MHz OH transition for ULIRGs in the literature. We interpret this as evidence that the OH abundances in both OHMs and non-masing galaxies are similar, and are not a limiting factor for megamaser emission. 

A companion paper (Willett et~al. 2011; Paper II) presents a full analysis of the mid-IR data, with comparisons between the two samples and connections to the OH megamaser properties. 


\acknowledgments

This work is based on observations made with the Spitzer Space Telescope, which is operated by the Jet Propulsion Laboratory, California Institute of Technology under a contract with NASA. Support for this work was provided by NASA through grant 30407 issued by JPL/Caltech. We have made extensive use of the NASA/IPAC Extragalactic Database (NED) which is operated by JPL and Caltech under contract with NASA. Many thanks are due to J.-D.~Smith for help with PAHFIT, N.~Halverson for comments on computing flux limits, D.~Farrah for useful discussions, and to the Spitzer Science Center for hosting KWW and JD in April 2007. VC acknowledges partial support from the EU ToK grant 39965 and FP7-REGPOT 206469.


\appendix

\section{High-resolution data with background sky subtraction}

As discussed in \S\ref{sec-obs}, the reduction process for the overall sample is slightly different for some archival galaxies that did not have separate IRS sky backgrounds in the high-resolution modules. Since much of our subsequent analysis (Paper II) depends on statistical comparisons between the two samples, we chose to minimize possible systematic errors and reduced all galaxies in a uniform manner {\em without} HR sky subtraction. These data are, however, likely to be a more reliable indicator of the absolute flux levels due to subtraction of the zodiacal background; therefore, we also present atomic and molecular line fluxes for these galaxies. 

\begin{deluxetable*}{lcccc}
\tabletypesize{\scriptsize}
\tablecaption{Hi-res line fluxes for common atomic emission lines with HR sky subtraction \label{tbl-atomicsky}}
\tablewidth{0pt}
\tablehead{
\colhead{Object} & 
\colhead{[S \tiny{IV}\scriptsize]} &
\colhead{[Ne \tiny{II}\scriptsize]} &
\colhead{[Ne \tiny{III}\scriptsize]} & 
\colhead{[S \tiny{III}\scriptsize]}
\\
\colhead{} & 
\colhead{10.511~\um} &
\colhead{12.814~\um} &
\colhead{15.555~\um} &
\colhead{18.713~\um}
}
\startdata
IRAS 01562+2528   &  --  &  --   &  --   & 0.65  \\
IRAS 02524+2046   &  --  &  --   &  --   & 0.47  \\
IRAS 04121+0223   &  --  &  --   &  --   & 0.67  \\
IRAS 04454$-$4838 & 0.42 & 1.95  & 0.43  & --    \\
IRAS 06487+2208   &  --  &  --   &  --   & 5.01  \\
IRAS 07163+0817   &  --  &  --   &  --   & 1.34  \\
IRAS 08201+2801   &  --  &  --   &  --   & 0.41  \\
IRAS 08449+2332   &  --  &  --   &  --   & 1.60  \\
IRAS 08474+1813   &  --  &  --   &  --   & 0.14  \\
IRAS 10035+2740   &  --  &  --   &  --   & 0.35  \\
IRAS 10039$-$3338 & 0.98 & 17.22 & 4.20  & 8.08  \\
IRAS 10339+1548   &  --  &  --   &  --   & 0.97  \\
IRAS 11180+1623   &  --  &  --   &  --   & 0.40  \\
IRAS 11524+1058   &  --  &  --   &  --   & 0.27  \\
IRAS 12540+5708   &  --  & 19.47 &  --   & --    \\
IRAS 14059+2000   &  --  &  --   &  --   & 0.59  \\
IRAS 14553+1245   &  --  &  --   &  --   & 1.58  \\
IRAS 15327+2340   &  --  & 59.39 & 6.73  & 7.54  \\
IRAS 16255+2801   &  --  &  --   &  --   & 1.54  \\
IRAS 18368+3549   &  --  &  --   &  --   & 1.01  \\
IRAS 18588+3517   &  --  &  --   &  --   & 2.89  \\
IRAS 20286+1846   &  --  &  --   &  --   & 0.48  \\
IRAS 21077+3358   &  --  &  --   &  --   & 0.74  \\
IRAS 21272+2514   &  --  & 2.22  & 0.33  & 0.40  \\
IRAS 22055+3024   &  --  &  --   &  --   & 1.16  \\
IRAS 22116+0437   &  --  &  --   &  --   & 0.90  \\
\hline
IRAS 00163$-$1039 & 2.64 & 87.43 & 14.30 & 32.10 \\
IRAS 05083+7936   & --   & 49.95 & 7.63  & 19.60 \\
IRAS 06538+4628   & 0.84 & 47.34 & 5.90  & 20.49 \\
IRAS 09437+0317   & --   & 8.72  & 1.22  & 3.58  \\
IRAS 23394$-$0353 & --   & 46.45 & 7.60  & 18.50 \\
\enddata
\tablecomments{Fluxes are in $10^{-21}$~W~cm$^{-2}$. No data is given for the 26 OHMs in our program for lines with $\lambda_{rest}\lesssim16$~\um~since we have no SH sky subtraction available for these objects.}
\end{deluxetable*}

\begin{deluxetable*}{lrrrrrrrrr}
\tabletypesize{\scriptsize}
\tablecaption{Hi-res line fluxes for rarer atomic emission lines with HR sky subtraction \label{tbl-unussky}}
\tablewidth{0pt}
\tablehead{
\colhead{Object} & 
\colhead{H \tiny{I} 7-6} &
\colhead{[Ne \tiny{V}\scriptsize]} & 
\colhead{[Cl \tiny{II}\scriptsize]} &
\colhead{[Fe \tiny{II}\scriptsize]} &
\colhead{[Ne \tiny{V}\scriptsize]} & 
\colhead{[O \tiny{IV}\scriptsize]} &
\colhead{[Fe \tiny{II}\scriptsize]} &
\colhead{[S \tiny{III}\scriptsize]} & 
\colhead{[Si \tiny{II}\scriptsize]}
\\
\colhead{$\lambda_{rest}$ [\um]} & 
\colhead{12.368} &
\colhead{14.322} &
\colhead{14.369} &
\colhead{17.936} &
\colhead{24.318} &
\colhead{25.890} &
\colhead{25.988} &
\colhead{33.481} &
\colhead{34.815}
}
\startdata
IRAS 04454$-$4838 & --   &  --  &  --  & --   & --   & 1.12 & --   & 0.90  & --    \\
IRAS 06487+2208   & --   &  --  &  --  & --   & --   & 1.12 & --   & --    & --    \\
IRAS 07163+0817   & --   &  --  &  --  & 0.73 & --   & --   & --   & --    & --    \\
IRAS 10339+1548   & --   &  --  &  --  & --   & 0.84 & 2.55 & --   & --    & --    \\
IRAS 11524+1058   & --   &  --  &  --  & 0.35 & --   & 0.59 & --   & --    & --    \\
IRAS 14553+1245   & --   &  --  &  --  & 0.41 & --   & --   & --   & --    & --    \\
IRAS 16255+2801   & --   &  --  &  --  & --   & --   & 0.46 & --   & --    & --    \\
IRAS 18368+3549   & --   &  --  &  --  & 0.51 & --   & --   & --   & --    & --    \\
IRAS 21272+2514   & 0.20 & 0.10 &  --  & --   & --   & --   & --   & --    & --    \\
\hline
IRAS 00163$-$1039 & 0.64 & --   & --   & --   & --   & 1.56 & 2.46 & 32.18 & 74.28 \\
IRAS 05083+7936   & --   & 0.58 & 0.43 & --   & --   & 1.53 & 1.97 & 29.34 & 43.28 \\
IRAS 06538+4628   & --   & 1.08 & --   & 1.06 & --   & --   & 2.15 & 37.21 & 48.72 \\
IRAS 09437+0317   & --   & --   & --   & --   & --   & 0.47 & 0.68 &  9.60 & 22.74 \\
IRAS 23394$-$0353 & 0.59 & --   & --   & --   & --   & 1.49 & 2.53 & 42.37 & 53.56 \\
\enddata
\tablecomments{Fluxes are in $10^{-21}$~W~cm$^{-2}$. No data is given for the 26 OHMs in our program for lines with $\lambda_{rest}\lesssim16$~\um~since we have no SH sky subtraction available for these objects.}
\end{deluxetable*}

\begin{deluxetable*}{lrrrr}
\tabletypesize{\scriptsize}
\tablecaption{Hi-res line fluxes and upper limits for H$_2$ transitions with HR sky subtraction \label{tbl-h2sky}}
\tablewidth{0pt}
\tablehead{
\colhead{Object} & 
\colhead{\htwo~S(3)} & 
\colhead{\htwo~S(2)} & 
\colhead{\htwo~S(1)} & 
\colhead{\htwo~S(0)}
\\
\colhead{$\lambda_{rest}$ [\um]} & 
\colhead{ 9.67} & 
\colhead{12.28} & 
\colhead{17.04} & 
\colhead{28.22}
}
\startdata
IRAS 01562+2528   &  --   &  --  & 0.78  & --    \\
IRAS 02524+2046   &  --   &  --  & 0.72  & --    \\
IRAS 04454$-$4838 &  1.05 & 1.21 & 3.03  & 0.76  \\
IRAS 06487+2208   &  --   &  --  & 2.03  & --    \\
IRAS 08201+2801   &  --   &  --  & 0.51  & --    \\
IRAS 08449+2332   &  --   &  --  & 1.40  & --    \\
IRAS 08474+1813   &  --   &  --  & 0.27  & --    \\
IRAS 10035+2740   &  --   &  --  & 1.16  & --    \\
IRAS 10039$-$3338 &  3.39 & 1.79 & 3.85  & --    \\
IRAS 10339+1548   &  --   &  --  & 0.46  & --    \\
IRAS 11180+1623   &  --   &  --  & 0.89  & --    \\
IRAS 11524+1058   &  --   &  --  & 0.81  & --    \\
IRAS 12540+5708   &  2.42 & 4.22 & --    & --    \\
IRAS 14059+2000   &  --   &  --  & 2.55  & --    \\
IRAS 15327+2340   &  --   & 7.36 & 15.42 & 14.55 \\
IRAS 16255+2801   &  --   &  --  & 0.55  & --    \\
IRAS 21077+3358   &  --   &  --  & 1.17  & --    \\
IRAS 21272+2514   &  0.26 & 0.35 & 0.68  & --    \\
IRAS 22116+0437   &  --   &  --  & 1.35  & --    \\
IRAS 23028+0725   &  --   &  --  & 1.11  & --    \\
\hline
IRAS 00163$-$1039 &  2.18 & 2.82 & 6.01  & --    \\
IRAS 05083+7936   &  2.34 & 2.64 & 5.11  & --    \\
IRAS 06538+4628   &  --   & 3.33 & 8.70  & 2.99  \\
IRAS 09437+0317   &  --   & --   & 2.67  & 1.78  \\
IRAS 23394$-$0353 &  3.40 & 2.25 & 4.99  & 1.38  \\
\enddata
\tablecomments{Fluxes are in $10^{-21}$~W~cm$^{-2}$. No data is given for the 26 OHMs in our program for the S(2) or S(3) lines since we have no SH sky subtraction available for these objects.}
\end{deluxetable*}


\bibliography{kwrefs}

\end{document}